\documentclass[10pt,onecolumn] {IEEEtran}

\usepackage{graphicx}
\usepackage{float}
\usepackage{amsthm}
\usepackage{amsmath}
\usepackage{amsfonts}
\usepackage{listings}
\usepackage{paralist}
\usepackage[usenames,dvipsnames]{xcolor}
\usepackage{cite}

\theoremstyle{plain}\newtheorem{thm}{Theorem}
\theoremstyle{definition}\newtheorem{defn}{Definition}
\theoremstyle{plain} \newtheorem{lemma}{Lemma}
\theoremstyle{plain} \newtheorem{cor}{Corollary}
\theoremstyle{remark}

\newcommand{\Frac}[2]{{{#1}/{#2}}}  
\DeclareMathOperator{\E}{E}
\DeclareMathOperator{\Prb}{P}
\def\APPROX{\simeq}
\newcommand{\beq}{\begin{equation}}
\newcommand{\eeq}{\end{equation}}

\newcommand{\Q}{Q_K}  

\newcommand{\Qc}{Q_{K,\lambda}}  
\newcommand{\QcV}{Q_{K_1^N,\lambda_1^N}}
  
\newcommand{\Lv}{\lambda_1^N}  
\newcommand{\Kv}{K_1^N}  
\newcommand{\Rv}{R_1^N} 

\newcommand{\Dmse}{D_\mathrm{mse}(K,\lambda)}

\newcommand{\DmseFRopt}{D^*_\mathrm{mse,fr}(R)}
\newcommand{\DmseECopt}{D^*_\mathrm{mse,ec}(R)}
\newcommand{\LmseFR}{\lambda_\mathrm{mse,fr}(x)}

\newcommand{\DfmseV}{D_\mathrm{fmse}(\Kv,\Lv)}

\newcommand{\DfmseFRopt}{D^*_\mathrm{fmse,fr}(\Rv)}
\newcommand{\DfmseECopt}{D^*_\mathrm{fmse,ec}(\Rv)}

\newcommand{\LfmseFRopt}{\lambda^*_{n,\mathrm{fmse,fr}}(x)}
\newcommand{\LfmseECopt}{\lambda^*_{n,\mathrm{fmse,ec}}(x)}

\usepackage{color}
\usepackage[normalem]{ulem} 

\begin{document}

\title{Distributed Quantization Networks}

\author{John~Z.~Sun,~\IEEEmembership{Student~Member,~IEEE,}
        and Vivek~K~Goyal,~\IEEEmembership{Senior~Member,~IEEE}
\thanks{J. Z. Sun is with the Department of Electrical Engineering and Computer
Science and the Research Laboratory of Electronics, Massachusetts Institute of
Technology, Cambridge, MA 02139 USA (e-mail: johnsun@mit.edu).}
\thanks{V. K. Goyal is with the Research Laboratory of Electronics,
Massachusetts Institute of Technology, Cambridge, MA 02139 USA
(e-mail: v.goyal@ieee.org).}
\thanks{This material is based upon work supported by the National Science Foundation under Grant No.\ 1115159.}}

\maketitle

\begin{abstract}
	Several key results in distributed source coding offer the intuition that little improvement in compression
	can be gained from intersensor communication when the information is coded in long blocks. 
	However, when sensors are restricted to code their observations in small blocks (e.g., 1), 
	intelligent collaboration between sensors can greatly reduce distortion.
	For networks where sensors are allowed to ``chat'' using a side channel that is unobservable at the fusion 
	center, we provide asymptotically-exact characterization of distortion performance and optimal 
	quantizer design in the high-resolution (low-distortion) regime using a framework called distributed 
	functional scalar quantization (DFSQ). 
	The key result is that chatting can dramatically improve performance even when intersensor communication 
	is at very low rate, especially if the fusion center desires fidelity of a nonlinear computation applied to 
	source realizations rather than fidelity in representing the sources themselves.
	We also solve the rate allocation problem when communication links have heterogeneous costs
	and provide a detailed example to demonstrate the theoretical and practical gains from chatting.
	This example for maximum computation gives insight on the gap between chatting and distributed networks,
	and how to optimize the intersensor communication.
\end{abstract}

\begin{IEEEkeywords}
distributed source coding,
high-resolution quantization,
sensor networks,
side information
\end{IEEEkeywords}

\section{Introduction}
\label{sec:intro}

A longstanding consideration in distributed compression systems is whether sensors wishing to convey information
to a fusion center should communicate with each other to improve efficiency.
Architectures that only allow communication between individual sensors and the fusion center 
simplify the network's communication protocol and decrease sensor responsibilities. 
Moreover, information theoretic results such as the Slepian--Wolf theorem show that distributed compression 
can perform as well as joint compression for lossless communication of correlated 
information sources~\cite{SlepianW:73}. 
Although this surprising and beautiful result does not extend fully, comparable results for lossy coding
show that the rate loss from separate encoding can be small using Berger--Tung coding
(see, e.g., \cite{Zamir:96}), again suggesting that communication between sensors has little or no utility. 

\begin{figure}
  \begin{center}
    \includegraphics[width=4in]{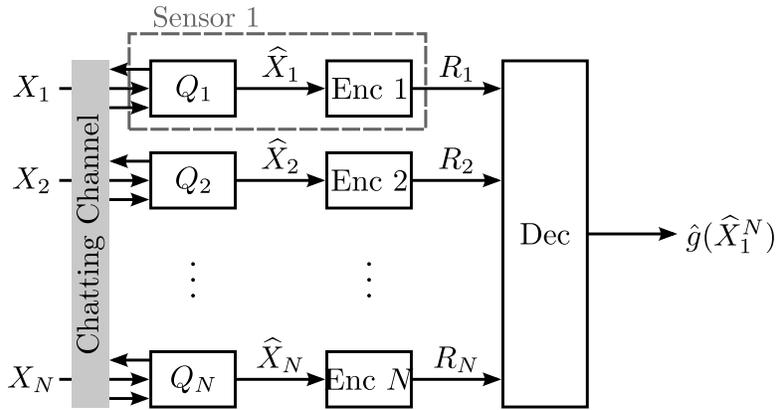}
  \end{center}
  \caption{A distributed computation network, where $N$ sensors (comprising quantizer and encoder)
  		observe realizations of correlated sources.
			Each observation $X_n$ is encoded and communicated over rate-limited links to a fusion center. 
			Simultaneously, each sensor can interact with a subset of other sensors using a noiseless but
			rate-limited chatting channel to improve compressibility.
			The decoder at the fusion center computes an estimate of the function 
			$g(X_1^n) = g(X_1,\,X_2,\,\ldots,\,X_n)$ 
			from the received data using a reconstruction function $\hat{g}(\widehat{X}_1^n)$
			but cannot observe messages communicated on the chatting channel. }
  \label{fig:star}
\end{figure}

Although it is tempting to use results from information theory to justify simple communication
topologies,
it is important to note the Slepian--Wolf result is dependent on large blocklength;
in the finite-blocklength regime, the optimality of distributed encoding does not hold~\cite{TanK:12arxiv}.
This paper examines the use of communication among sensors when the compression blocklength is 1, a regime where collaboration, called \emph{chatting} in this work, can greatly decrease 
the aggregate communication from sensors to the fusion center to meet a distortion criterion
as compared to a distributed network.
We analyze chatting networks using the distributed functional scalar quantization (DFSQ) framework,
which constrains sensors to using scalar quantizers to compress their observations and generalizes the fusion
center's objective to desire fidelity in computing a function of the sources rather than determining
the sources themselves~\cite{MisraGV:11,SunMG:12arxiv}.
Our problem model is shown in Fig.~\ref{fig:star},
where $N$ correlated but memoryless continuous-valued, discrete-time stochastic processes
produce scalar realizations $X_1^N(t) = (X_1(t) , \ldots , X_N(t))$ for $t \in \mathbb{Z}$.
For each $t$, realizations of these sources are scalar quantized by sensors 
and transmitted to a fusion center at rates $R_1^N$.
To aid this communication,
sensors can collaborate with each other via a side channel that is 
unobservable to the fusion center.
Since the quantization is scalar and the sources are memoryless, we remove the time index 
and model the sources as being drawn from a joint distribution $f_{X_1^N}$ at each $t$.

The side channel facilitating intersensor communication has practical implications.  
In typical communication systems, the transmission power needed for reliable communication
increases superlinearly with distance and bandwidth~\cite{TseV:05}. 
Hence, it is much cheaper to design short and low-rate links between sensors
than reliable and high-rate links to a fusion center. 
Moreover, milder transmission requirements provide more flexibility in determining the transmission media
or communication modalities employed, which can allow intersensor communication to be orthogonal to 
the main network. 
One such example is cognitive radio, a paradigm where the wireless spectrum can have secondary users that
communicate only when the primary users are silent~\cite{YucekA:09}.  
This means secondary users have less priority and hence
lower reliability and rate, which is adequate for intersensor communication. 

The main contributions of the paper are to precisely characterize the distortion
performance of a 
distributed network when chatting is allowed and to identify the optimal quantizer design for each sensor. 
We show that collaboration can have significant impact on performance;
in some cases, it can dramatically reduce distortion even when the chatting has extremely low rate. 
We also give necessary conditions on the chatting topology and protocol for successful decodability 
in the DFSQ framework, thus providing insight into the architecture design for chatting networks.
Finally, we recognize that intersensor communication can occur on low-cost channels and solve the rate allocation
problem in networks with heterogeneous links and different costs of transmission.
The basic concepts of this work were introduced in~\cite{SunG:12}; this paper provides more complete 
and definitive coverage, including more results on rate allocation, a discussion on 
generalizing chatting messages, and details on the impact of various optimizations.

We begin by introducing related work, notation and prerequisite results in Section~\ref{sec:prelim}. 
In Section~\ref{sec:perf}, we analyze the performance of chatting networks and discuss how
to optimize the communication that occurs.   
We then determine the proper rate allocation for chatting networks in Section~\ref{sec:chatall}. 
Finally, we develop intuition for the behavior of chatting by considering a maximum computation
network in Section~\ref{sec:max}; this specific example demonstrates the incremental gains achieved by
incorporating the different optimizations discussed in the paper.

\section{Preliminaries}
\label{sec:prelim}

\subsection{Previous Work}
\label{sec:prelim:previous}

There is a large body of literature studying asymptotic performance of the distributed network
in Fig.~\ref{fig:star} without the chatting channel; a comprehensive review of these works and their 
connections to DFSQ appears in~\cite{MisraGV:11}. 
Similarly, connections to coding for computing (e.g.\ \cite{OrlitskyR:01,FengES:04}) are discussed
there as well. 
Recent work on the finite-blocklength regime~\cite{PolyanskiyPV:10} has led to extensions in source
coding~\cite{IngberK:11,KostinaV:12,TanK:12arxiv}. 
In general, this analysis technique is meaningful for blocklengths as low as 100, but is unsuitable for regimes
traditionally considered in high-resolution theory. 

We review results that relate to the chatting channel, focusing on Shannon-theoretic results.
Kaspi and Berger provided inner bounds for the rate region of a two-encoder problem where one encoder can
send information for the other using compress-and-forward techniques~\cite{KaspiB:82}.
Recently, this bound has been generalized in~\cite{SefidgaranT:12}, 
but the exact rate region is still unknown except in special cases.
Chatting is related to source coding problems such as interaction~\cite{NaI:11,NaIG:12},
omniscience~\cite{NitinawaratN:10} and data exchange~\cite{CourtadeW:10}. 
However, these settings are more naturally suited for discrete-alphabet sources and existing results rely 
on large-blocklength analysis. 

There are also strong connections between this work and distortion side information~\cite{MartinianWZ:08}
and vector quantization with alternative distortion measures~\cite{LinderZZ:99}.

\subsection{Quantization}
\label{sec:prelim:quant}

The focus of this work is on compression of continuous-valued, finite-support sources 
using small blocks of data.
Here, performance results from Shannon theory are overly optimistic since tools such as joint-typicality
encoding and decoding are not reliable without operating far from the distortion--rate bound.
Instead, we consider the complementary asymptotic of high resolution, 
where the blocklength is small and the compression rate $R$ is large~\cite{GershoG:92,Neuhoff:93,GrayN:98}.
Before introducing the high-resolution asymptotic, we summarize the quantization model for the case of
blocklength 1 and set up the notation used for the rest of the paper.

A scalar quantizer $Q_K$ is a mapping from the real line to a set of $K$ points
$\mathcal{C} = \{c_k\}_{k=1}^K \subset \mathbb{R}$ called the codebook,
where $Q_K(x) = c_k$ if $x \in P_k$ and the cells
$\{ P_k \}_{k=1}^K$ form a partition of $\mathbb{R}$. 
The quantizer is called \emph{regular} if the partition cells are intervals containing the corresponding codewords.
For simplicity, the codebook and partition are indexed from smallest to largest,
implying $p_0 < c_1 \leq p_1 < c_2 \leq \cdots < c_K \leq p_K$ if $P_k = (p_{k-1}, p_k]$,
with $p_0 = -\infty$ and $p_K = \infty$.
Define the \emph{granular} region as $(c_1,c_K)$ and its complement
$(-\infty,c_1] \cup [c_K,\infty)$ as the \emph{overload} region. 

\begin{figure}
  \begin{center}
    \includegraphics[width=3.3in]{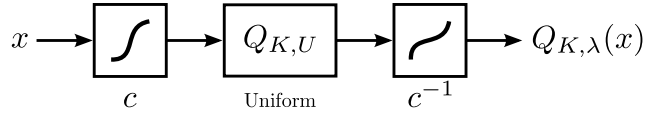}
  \end{center}
  \caption{The companding model is a method to construct nonuniform quantizers using a monotonic
  		nonlinearity $c$ satisfying $\lim_{x\rightarrow-\infty} c(x) = 0$ and $\lim_{x\rightarrow\infty} c(x) = 1$.
  		The notation $Q_{K,U}$ is used to describe the canonical uniform quantizer with $K$ codewords in the 
  		granular region $[0,1]$.}
  \label{fig:compander}
\end{figure}

Uniform quantization, where partition cells in the granular region have equal length,
is most common in practice, but nonuniform quantization can be better for compression if
the source can be modeled properly.
One way of constructing a nonuniform quantizer is using the compander model, 
where the scalar source is transformed using a nondecreasing and smooth
\emph{compressor} function $c : \mathbb{R} \rightarrow [0,1]$, 
then quantized using a uniform quantizer comprising $K$ levels on the granular region $[0,1]$, 
and finally passed through the \emph{expander} function $c^{-1}$ (Fig.~\ref{fig:compander}).
Compressor functions are defined such that $\lim_{x\rightarrow-\infty} c(x) = 0$ and
$\lim_{x\rightarrow\infty} c(x) = 1$.
It is convenient to define a \emph{point density function} as $\lambda(x) = c^\prime(x)$.
Because of the boundary conditions on $c$, there is a one-to-one correspondence between $\lambda$ and $c$;
hence, a companding quantizer can be uniquely specified using a point density function and codebook size, 
and is denoted $\Qc$ in this work.
The conversion of point density functions to finite-codeword quantizers is described in more detail in~\cite[Section~II-B]{SunMG:12arxiv}.

\subsection{High-resolution Theory}
\label{sec:prelim:hr}

It is generally difficult to determine the distortion of a scalar quantizer for any codebook size $K$.
However, the performance of $\Qc$ can be precisely analyzed as the number of codewords $K$ becomes large,
which is the basis of high-resolution theory. 
Assume a source $X$ is a continuous random variable, and define the mean squared error (MSE) distortion as
\beq
	\label{eq:mse:eq}
	\Dmse = \E [ |X - \Qc(X)|^2 ] ,
\eeq
where the expectation is with respect to the source density $f_X$.
Under the additional assumption that the tails of $f_X$ decay sufficiently fast,
\beq
	\label{eq:mse:approx}
	\Dmse \APPROX \frac{1}{12 K^2} \E [ \lambda^{-2}(X) ] ,
\eeq
where $\APPROX$ indicates that the ratio of the two expressions approaches 1 
as $K$ increases~\cite{Bennett:48,PanterD:51}.
Hence, the MSE performance of a scalar quantizer can be approximated by a simple
relationship between the source distribution, point density, and codebook size, and this
relation becomes more precise with increasing $K$. 
In fact, companding quantizers are \emph{asymptotically optimal}, meaning that the quantizer optimized 
over $\lambda$ has distortion that approaches the performance of the best $\Q$
found by any means~\cite{BucklewW:82,CambanisG:83,Linder:91}.
Experimentally, the high-resolution approximation is accurate even for moderate $K$~\cite{Neuhoff:93,Goyal:00b}. 

When the quantized values are to be communicated or stored, it is natural to map each codeword to 
a string of bits and consider the trade-off between performance and communication rate $R$, 
defined to be the expected number of bits per sample.
In the simplest case, the codewords are indexed with equal-length labels and the communication rate is $R = \log_2(K)$; 
this is called \emph{fixed-rate} or \emph{codebook-constrained} quantization. 
Since the distortion's dependence on the shape of the quantizer $\lambda$ is explicit in the asymptote, 
calculus techniques can be used to optimize companders.
For fixed rate, H\"{o}lder's inequality can show the optimal point density satisfies
\beq
	\label{eq:mse:lamfr}
	\LmseFR \propto f_X^{1/3}(x) ,
\eeq
and the resulting distortion is 
\beq
	\label{eq:mse:fr:distopt}
	\DmseFRopt \APPROX \frac{1}{12} \, \| f_X \|_{1/3} \, 2^{-2R},
\eeq
with the notation $\|f\|_p = (\int_{-\infty}^\infty f^p(x) \, dx)^{1/p}$~\cite{GrayG:77}.
The limit conditions on $c(x)$ imply the integral of $\lambda(x)$ is unity.
Thus, \eqref{eq:mse:lamfr} specifies the point density uniquely;
for clarity, we omit the normalization when presenting point density results. 

In general, the codeword indices can be coded to produce bit strings of different lengths based
on probabilities of occurrence;
this is referred to as \emph{variable-rate} quantization.
If the decoding latency is allowed to be large, one can employ block entropy coding and the 
communication rate approaches $H(\Qc(X))$.
This particular scenario, called \emph{entropy-constrained} quantization, can be analyzed using Jensen's
inequality to show the optimal point density $\lambda^*_\mathrm{mse,ec}$ is constant 
on the support of the input distribution~\cite{GrayG:77}.
The optimal quantizer is thus uniform, and the resulting distortion is
\beq
	\label{eq:mse:vr:distopt}
	\DmseECopt \APPROX \frac{1}{12} 2^{-2(R-h(X))} .
\eeq
Note that block entropy coding suggests that the sources are transmitted in blocks even though 
the quantization is scalar.
As such, \eqref{eq:mse:vr:distopt} is an asymptotic result and serves as a lower bound on practical entropy 
coders with finite blocklengths that match the latency restrictions of a network.

\subsection{Distributed Functional Scalar Quantization}
\label{sec:prelim:dfsq}

When the goal of acquisition is to approximate some computation applied to the sources, 
optimizing the compression to the source distribution can be suboptimal and potentially
worse than uniform quantization. 
This is most evident in distributed networks since each sensor cannot determine the overall computation
at the encoder.
The distributed functional scalar quantization (DFSQ) framework accounts for the computational task at the 
fusion center, and the resulting quantizers can be substantially better than
naive designs~\cite{MisraGV:11,SunMG:12arxiv}. 
In this setting, the distortion criterion is functional MSE (fMSE):
\beq
	\label{eq:fmse} 
	\DfmseV= \E \left[ \big|g(X_1^N) - \hat{g}(\QcV(X_1^N)) \big|^2 \right] ,
\eeq
where $g$ is a scalar function of interest, $\hat{g}$ is the decoding function
and $\QcV$ is scalar quantization performed on a vector such that
\[ \QcV(x_1^N) = \left( Q_{K_1, \lambda_1}(x_1) , \ldots Q_{K_N, \lambda_N}(x_N) \right) \mbox{.} \]

Before understanding how a quantizer changes fMSE, it is convenient to define how a computation locally 
affects distortion. 
\begin{defn}[\!\!\cite{MisraGV:11}]
	\label{def:sens}
	The $n$th \emph{functional sensitivity profile} of a multivariate function $g$ is defined as 
	\beq
		\label{eq:gamman}
		\gamma_n(x) = \left( \E \left[ |g_n(X_1^N)|^2 \, \middle| \, X_n = x \right]\right)^{1/2} ,
	\eeq
	where $g_n(x)$ is the partial derivative of $g$ with respect to its $n$th argument evaluated at the point $x$.
\end{defn}

Given the sensitivity profile, the main result of DFSQ~\cite{MisraGV:11} says the distortion
of a set of $N$ companding quantizers has the asymptotic form
\beq
	\label{eq:fmse:dist} 
	\DfmseV \APPROX \sum_{n=1}^N \frac{1}{12 K_n^2} 
					\E \left[ \left( \frac{\gamma_n(X_n)}{\lambda_n(X_n)} \right)^2 \right] ,
\eeq
with conditional expectation decoder 
\beq
	\label{eq:optdec}
	\hat{g}(x_1^N) = \E \left[ g(X_1^N) \, \middle| \, \QcV(X_1^N) = \QcV(x_1^N) \right] ,
\eeq
provided the following conditions are satisfied:
\begin{asparaenum}[MF1.]
	\item The function $g$ is Lipschitz continuous and twice differentiable in every argument
					except possibly on a set of Jordan measure 0. 
	\item The source pdf $f_{X_1^N}$ is continuous, bounded, and supported on $[0,1]^N$.
	\item The function $g$ and set of point densities $\lambda_1^N$ allow
					$\E[( \Frac{\gamma_n(X_n)}{\lambda_n(X_n)} )^2 ]$ to be defined and finite for all $n$. 
\end{asparaenum}
Similar conditions are given in~\cite{SunMG:12arxiv} for infinite-support distributions and a simpler decoder. 

Following the same recipes to optimize over $\Lv$ as in the MSE setting, the relationship between
distortion and communication rate is found.
In both cases, the sensitivity acts to shift quantization points to where they can reduce the 
distortion in the computation. 
For fixed-rate quantization, the asymptotic minimum distortion is
\beq
	\label{eq:dfsq:distfr}
	\DfmseFRopt \APPROX \sum_{n=1}^N \frac{1}{12} \, \| \gamma_n f_{X_n} \|_{1/3} \, 2^{-2R_n} ,
\eeq
where $f_{X_n}$ is the marginal distribution of $X_n$ and each optimal point density satisfies
\beq
	\label{eq:dfsq:lamfr}
	\LfmseFRopt \propto \left( \gamma_n(x) f_{X_n}(x) \right)^{1/3} .
\eeq
Meanwhile, for entropy-constrained quantization,
the asymptotic minimum distortion is 
\beq
	\label{eq:dfsq:distec}
	\DfmseECopt \APPROX \sum_{n=1}^N \frac{1}{12} 2^{2 h(X_n) + 2 \E\left[ \log_2 \gamma(X_n) \right]} 2^{-2R_n} ,
\eeq
which results from point densities satisfying
\beq
	\label{eq:dfsq:lamec}
	\LfmseECopt \propto \gamma_n(x) .
\eeq

\subsection{Don't-care intervals}

When the computation induces the sensitivity to be 0 on some subintervals of the support, the high-resolution
assumptions are violated and the asymptotic distortion performance may not be described by~\eqref{eq:fmse:dist}. 
This issue is addressed by carefully coding when the source is in such 
a ``don't-care'' interval~\cite[Section~VII]{MisraGV:11} and then applying traditional 
high-resolution theory to the remaining support. 
This consideration is particularly relevant because chatting among sensors can often induce the 
conditional sensitivity to be 0, and proper coding can lead to greatly improved performance. 

Consider $L_n$ don't-care intervals in $\gamma_n$ and let $A_n$ be the event that the source 
realization is \emph{not} in the unions of them.
In the fixed-rate setting, one codeword is allocated to each don't-care interval, and the remaining
$K_n - L_n$ codewords are used to form reconstruction points in the nonzero intervals. 
There is a small degradation in performance from the loss corresponding to $L_n$, but this quickly
becomes negligible as $K_n$ increases.
In the entropy-constrained case, 
the additional flexibility in coding allows for the encoder to split its message and reduce cost. 
The first part is an indicator variable revealing whether the source is in a don't-care interval 
and can be coded at rate $I_A \equiv H_B(\Prb(A_n))$, where $H_B$ is the binary entropy function.
The actual reconstruction message is only sent if event $A_n$ occurs,
and its rate is amplified to $(R_n- I_A)/\Prb(A_n)$ to meet the average rate constraint. 
The multiplicative factor $1/\Prb(A_n)$ is called the \emph{rate amplification}.

\subsection{Chatting}

In~\cite[Section~VIII]{MisraGV:11}, chatting is introduced in the setting where one sensor sends exactly one bit 
to another sensor. 
Under fixed-rate quantization, this collaboration can at most decrease the distortion by a factor of 4 using a 
property of $\mathcal{L}_{1/3}$ quasi-norms. 
Because utilizing that bit to send additional information to the fusion center would decrease distortion 
by exactly a factor of 4, this is considered a negative result.
Here, there is an implicit assumption that links have equal cost per bit and the network wishes to 
optimize a total cost budget. 
In the entropy-constrained setting, chatting may be useful even when links have equal costs. 
One example was given to demonstrate a single bit of chatting can decrease the 
distortion by an unbounded amount;
more generally, the benefit of chatting varies depending on the source joint distribution and 
decoder computation. 

In previous work, there is no systematic theory on performance and quantizer design of chatting.
Moreover, collaboration in larger networks was still an open problem. 
In this paper, we extend previous results and provide a more complete discussion on how a chatting channel
affects a distributed quantization network.
A sample result is that chatting can be beneficial in the fixed-rate setting if the cost of communicating
a bit to another sensor is lower than the cost of communicating a bit to the fusion center.

\section{Performance and Design of Chatting Networks}
\label{sec:perf}

We model the chatting channel in Fig.~\ref{fig:star} as a directed graph 
$\mathcal{G}^c = (\mathcal{V},\mathcal{E})$, where
the set of nodes $\mathcal{V}$ is the set of all sensors and 
$\mathcal{E} \subseteq \mathcal{V} \times \mathcal{V}$ is the set of noiseless,
directed chatting links. 
If $(i,n) \in \mathcal{E}$, then for each source realization,
Sensor~$i$ sends to Sensor~$n$ a chatting message
$M_{i \to n}$ with codebook size $K_{i \to n}$.
The parent and children sets of a sensor $n \in \mathcal{V}$ are denoted
$\mathcal{N}_p(n)$ and $\mathcal{N}_c(n)$ respectively;
when $(i,n) \in \mathcal{E}$,
$i$ is a parent of $n$ and $n$ is a child of $i$.
The set of all chatting messages is $M^c = \{M_{i \to n}\}_{(i,n) \in \mathcal{E}}$
and the set of corresponding codebook sizes is $K^c = \{K_{i \to n}\}_{(i,n) \in \mathcal{E}}$. 
The chatting messages are communicated according to
a schedule that the sensors and the fusion center know in advance;
the set of chatting messages $M^c$ can therefore also be thought of as a sequence.
We assume chatting occurs quickly in that all communication is completed
before the next discrete time instant (at which point new realizations of
$X_1^N$ are measured).
After chatting is complete, Sensor~$n$ compresses its observation
$X_n$ into a message $M_n$ using a codebook dependent on the information
gathered from chatting messages, which is noiselessly communicated to the fusion center 
with a message $M_n(M^c)$ with codebook size $K_n(M^c)$.

We now present fMSE performance of $\QcV$ in the fixed-rate and entropy-constrained settings,
and we show how to optimize $\Lv$ given $K_1^N$ and $K^c$. 
We first analyze the network assuming the fusion center can successfully 
infer the codebook used by each sensor and hence recover the quantized values from messages $M_1^N$.
Later in Section~\ref{sec:perf:cond}, we provide conditions on the chatting graph $\mathcal{G}^c$ and 
set of chatting messages $M^c$ such that the fusion center is successful with zero error, 
having benefited from already understanding the quantizer design. 

Before studying fMSE, we need to extend the definition of functional sensitivity.
\begin{defn}
	Let $\mathcal{N}_p(n) \subseteq V$ be the set of parents of Sensor~$n$ in the graph $\mathcal{G}^c$ 
	induced by chatting.
	The $n$th \emph{conditional sensitivity profile} of computation $g$ given all chatting messages $M^c$ is
	\beq
		\label{eq:condsens}
		\gamma_{n|M^c}(x|m) = \left( \E \left[ | g_n(X_1^N) |^2 \, \middle| \,  X_n = x, 
									M_{i \to n}  = m_{i \to n} \ \mbox{for all} \ i \in \mathcal{N}_p(n) 
											\right] \right)^{1/2} . 
	\eeq
\end{defn}
Notice only messages from parent sensors are relevant to $\gamma_{n|M^c}$.
Intuitively, chatting messages reveal information about the parent sensors' quantized values
and reshape the sensitivity appropriately. 
Depending on the encoding of chatting messages, 
this may induce don't-care intervals in the conditional sensitivity (where $\gamma_{n|M^c} = 0$). 

The distortion dependence on the number of codeword points and the conditional sensitivity profiles 
is given in the following theorem:

\begin{thm}
	\label{thm:chatdist}
	Given the source distribution $f_{X_1^N}$, computation $g$, and point densities $\Lv(M^c)$ satisfying
	conditions MF1--3 for every possible realization of $M^c$, the asymptotic distortion
	of the conditional expectation decoder~\eqref{eq:optdec}
	given codeword allocation $K_1^N$ and $K^c$ is
	\beq
	 	D_\mathrm{fmse}(\Kv, K^c, \Lv) \APPROX \E_{M^c} \left[ \sum_{n=1}^N \E_{X_n|M^c} 
	 		 			\left[ \frac{1}{12 K^2_n(m)} 
							\frac{\gamma^2_{n|M^c} (X_n|m)}{\lambda^2_{n|M^c} (X_n|m)} \ \middle| \ M^c = m \right] \right] . 
	\eeq
\end{thm}
\begin{IEEEproof}
	Extend the proof of~\cite[Theorem 17]{MisraGV:11} using the Law of Total Expectation.
\end{IEEEproof}

Compared to the DFSQ result, the performance of a chatting network can be substantially more difficult to 
compute since the conditional sensitivity may be different with each realization of $M^c$
and affects the choice of the point density and codebook size. 
However, Sensor~$n$'s dependence on $M^c$ is through a subset of messages from its parent nodes.  
In Section~\ref{sec:max}, we will see how structured architectures lead to tractable computations of fMSE\@.
Following the techniques in~\cite{SunMG:12arxiv},
the theorem can be expanded to account for infinite-support distributions and a simpler decoder.
Some effort is necessary to justify the use of normalized point densities in the infinite-support case, 
especially in the entropy-constrained setting, but high-resolution theory applies in this case as well.

\subsection{Don't-Care Intervals}
\label{sec:perf:dontcare}

We have already alluded to the fact that chatting can induce don't-care intervals in the conditional
sensitivity profiles of certain sensors. 
In this case, we must properly code for these intervals to ensure the high-resolution assumptions hold,
as discussed in Section~\ref{sec:prelim:dfsq}.

For fixed-rate coding where $R_n = \log_2(K_n)$, this means shifting one codeword to the interior of each 
don't-care interval and applying standard high-resolution analysis over the union of all intervals where $\gamma_n(x) > 0$.
The resulting distortion of a chatting network is then given as:

\begin{cor}
	\label{cor:fMSE:fr}
	Assume the source distribution $f_{X_1^N}$, computation $g$, and point densities $\Lv(M^c)$ satisfying
	conditions MF1--3 for every possible realization of $M^c$, with the additional requirement that
	$\lambda_n(x \,|\, m) = 0$ whenever $\gamma_{n|M^c}(x \,|\, m) = 0$.
	Let $L_n(m)$ be the number of don't-care intervals in the conditional sensitivity of Sensor~$n$ when 
	$M^c = m$.
	The asymptotic distortion of such a chatting network where communication links utilize fixed-rate coding is
	\beq
	 	D_\mathrm{fmse}(\Rv, K^c, \Lv) \APPROX \E_{M^c} \left[ \sum_{n=1}^N \E_{X_n|M^c} 
	 		 			\left[ \frac{1}{12 ( 2^{R_n} - L_n(m))} 
							\frac{\gamma^2_{n|M^c} (X_n \,|\, m)}{\lambda^2_{n|M^c} (X_n \,|\, m)}
							\ \middle| \ M^c = m \right] \right] . 
	\eeq
\end{cor}

In the entropy-constrained setting where $R_n = H(\widehat{X}_n)$, we must code first the event $A_n(m)$ that
the source is not in a don't-care interval given the chatting messages, and then coding
the source realization only if $A_n$ occurs. 
The resulting distortion of a chatting network is:

\begin{cor}
	\label{cor:fMSE:ec}
	Assume the source distribution $f_{X_1^N}$, computation $g$, and point densities $\Lv(M^c)$ satisfying
	conditions MF1--3 for every possible realization of $M^c$, with the additional requirement that
	$\lambda_n(x \,|\, m) = 0$ whenever $\gamma_{n|M^c}(x \,|\, m) = 0$.
	Let $A_n(m)$ be the event that $X_n$ is not in a don't-care interval given $M^c = m$.
	The asymptotic distortion of such a chatting network where communication links utilize entropy coding is
	\begin{align*}
	 	D_\mathrm{fmse}(\Rv, K^c, \Lv) &\APPROX \E_{M^c} \left[ \sum_{n=1}^N \E_{X_n|M^c} 
	 		 			\left[ \frac{\Prb(A_n(m))}{12} 2^{2 h(X_n | A_n(m)) + 2 \E [ \log_2 \lambda_n(X_n) | A_n(m)]} 
	 		 						\right. \right. \\
							& \hspace{12ex} \left. \left. 
										\cdot \frac{\gamma^2_{n|M^c} (X_n \,|\, m)}{\lambda^2_{n|M^c} (X_n \,|\, m)} 
										2^{-2 ( R_n(m) - H_B(A_n(m)))/P(A_n(m))} \ \middle| \ M^c = m \right] \right] . 
	\end{align*}
\end{cor}

We will use both corollaries in optimizing the design of $\Lv(M^c)$ in the remainder of the paper.

\subsection{Fixed-rate Quantization Design}
\label{sec:perf:fr}

We mirror the method used to determine~\eqref{eq:dfsq:lamfr} in the DFSQ setup
but now allow the sensor to choose from a set of codebooks depending on the incoming messages
from parent sensors. 
The mapping between chatting messages and codebooks is known to the decoder of the fusion center, 
and each codebook corresponds to the optimal quantizer for a given conditional sensitivity induced 
by the incoming message.
Let $Z_n(M^c)$ be the union of the don't-care intervals of a particular conditional sensitivity.
Then using Corollary~\ref{cor:fMSE:fr}, the optimal point density for fixed-rate quantization satisfies
\beq
	\label{eq:chat:lamfr}
	\lambda^*_{n,\mathrm{fmse,fr,chat}}(x \,|\, m) \propto \left\{
				\begin{array}{ll}
         	\left( \gamma_{n|M^c}(x \,|\, m) f_{X_n|M^c}(x \,|\, m) \right)^{1/3} , & 
         												x \notin Z_n(m) \mbox{ and } f_{X_n|M^c}(x \,|\, m) > 0 ; \\
       			0, & \mathrm{otherwise}.
     \end{array} 
   \right. 
\eeq

Recall that the point density is the derivative of the compressor function $c(x)$ in
the compander model. 
Hence, codewords are placed at the solutions to $c(x) = \Frac{(k-1)}{(K-L)}$ for $k = 1, \dots, (K-L)$.
In addition, one codeword must be placed in each of the $L$ don't-care interval.

\subsection{Entropy-constrained Quantization Design}
\label{sec:perf:ec}

Using Corollary~\ref{cor:fMSE:ec}, the optimal point density when entropy coding is combined with 
scalar quantization has the form
\beq
	\label{eq:chat:lamec}
	\lambda^*_{n,\mathrm{fmse,ec,chat}}(x \,|\, m) \propto \left\{
					\begin{array}{ll}
         		\gamma_{n|M^c}(x \,|\, m) , &  x \notin Z_n(m) \mbox{ and } f_{X_n|M^c}(x \,|\, m) > 0 ; \\
       			0, & \mathrm{otherwise}.
     \end{array}
   \right.
\eeq
Note that rate amplification can arise through chatting, and this can allow distortion terms to decay
at rates faster than $2^{-2R_n}$.
However, there is also a penalty from proper coding of don't-care intervals, corresponding to 
$H_B( P(A_n) )$. 
This loss is negligible in the high-resolution regime but may become important for moderate rates.

\subsection{Conditions on Chatting Graph}
\label{sec:perf:cond}

We have observed that chatting can influence optimal design of scalar quantizers through the 
conditional sensitivity, and that sensors will vary their quantization codebooks depending on the 
incoming messages from parent sensors.
Under the assumption that the fusion center does not have access to $M^c$, success of compression is 
contingent on the fusion center identifying the codebook employed by every sensor from the messages $M_1^N$.

\begin{defn}
	A chatting network is \emph{codebook identifiable} if the fusion center can determine the 
	codebooks of $\QcV$ using the messages it receives from each sensor. 
	That is, it can determine $\mathcal{C}_n(M^c)$ from $M_1^N$ for each time instant. 
\end{defn}

We have argued that a chatting network can successfully communicate its compressed observations if 
it is codebook identifiable. 
The following are sufficient conditions on the chatting graph $\mathcal{G}^c$ and messages $M^c$ 
such that the network is codebook identifiable:
\begin{asparaenum}[C1.]
	\item The chatting graph $\mathcal{G}^c$ is a directed acyclic graph. 
	\item The causality in the chatting schedule matches $\mathcal{G}^c$, meaning for every $n$,
						Sensor~$n$ sends its chatting message after it receives messages from from all parent sensors.
	\item The quantizer at Sensor~$n$ is a function of the source joint distribution and all 
						incoming messages from parent sensors in $\mathcal{N}_p(n)$. 
	\item At any discrete time, the message transmitted by Sensor~$n$ is a function of 
						$M_n$ and incoming messages from parent sensors in $\mathcal{N}_p(n)$.
\end{asparaenum}

When each sensor's quantizer is regular and encoder only operates on the quantized values $\widehat{X}_n$,
matching the DFSQ setup, the chatting message can only influence the choice of codebook.
In this setting, the above conditions become necessary as well.
Alternatively, if sensors can locally fuse messages from parents with their own observation, 
there may exist other conditions for a network to be codebook identifiable.

\section{Rate Allocation in Chatting Networks}
\label{sec:chatall}

A consequence of chatting is that certain sensors can exploit their neighbors' acquisitions 
to refine their own. 
Moreover, a sensor can potentially utilize this side information to adjust its communication rate 
in addition to changing its quantization if the network is codebook identifiable.
These features of chatting networks suggest intelligent rate allocation across sensors can yield significant 
performance gains. 
In addition, a strong motivation for intersensor interaction is that sensors may be geographically closer to 
each other than a fusion center and hence require less transmit power, or can utilize low-bandwidth orthogonal 
channels that do not interfere with the main communication network. 
As a result, the cost of communicating a bit may vary in a network.

This section explores proper rate allocation to minimize the total cost of transmission in a chatting network,
allowing asymmetry of the information content at each sensor and heterogeneity of the communication links.
Consider the distributed network in Fig.~\ref{fig:star}.
The cost per bit of the communication link and the resource allocation between Sensor~$n$ and the fusion center 
are denoted by $\alpha_n$ and $b_n$ respectively, leading to a communication rate of $R_n = b_n / \alpha_n$ 
from Sensor~$n$ to the fusion center.
Similarly, for a chatting link between Sensors $i$ and $n$, the cost per bit and resource allocation
are denoted by $\alpha_{i \to n}$ and $b_{i \to n}$ respectively, corresponding to a chatting rate of
$R_{i \to n} = b_{i \to n}/\alpha_{i \to n}$. 
Consistent with previous notation, we denote the set of costs per chatting bit,
resource allocations on chatting links, and chatting rates by
$\alpha^c = \{\alpha_{i \to n}\}_{(i,n)\in\mathcal{E}}$,
$b^c = \{b_{i \to n}\}_{(i,n)\in\mathcal{E}}$, and
$R^c = \{R_{i \to n}\}_{(i,n)\in\mathcal{E}}$.

Given a total resource budget $C$, how should the rates be allocated among these links?
For simplicity, assume all chatting links employ fixed-rate quantization;
this implies that $K_n = 2^{R_n}$
for all $n \in \{1,\,2,\,\ldots,\,N\}$
and $K_{i \to n} = 2^{R_{i \to n}}$
for all $(i,n) \in \mathcal{E}$.
The distortion--cost trade-off is then expressed as 
\beq
	D(C) = \inf_{\substack{b_1^N, b^c, \lambda_1^N : \\
							\sum_{n=1}^N b_n + \sum_{(i,n) \in \mathcal{E}} b_{i \to n} = C}}				 
							D_\mathrm{fmse} \left( K_1^N, K^c, \lambda_1^N \right) .
\eeq

In general, this optimization is extremely difficult to describe analytically since the distortion contribution
of each sensor is dependent in a nontrivial way on the conditional sensitivity, which in turn
is dependent on the design of the chatting messages. 
However, the relationship between $b_1^N$ and the overall system distortion is much simpler, as described
in Theorem~\ref{thm:chatdist}. 
Hence, once the chatting allocations $b^c$ is fixed, the optimal $b_1^N$ is easily determined using 
extensions of traditional rate allocation techniques described in Appendix~\ref{app:allocation}.
In particular, the optimal  $b_1^N$ can be found by applying Lemmas~\ref{lem:costop:lagrange} and 
\ref{lem:costop:lagrange_prob} with a total cost constraint
\beq
	C^\prime = C - \sum_{(i,n) \in \mathcal{E}} b_{i \to n} .
\eeq
A brute-force search over $b^c$ then provides the best allocation, but this procedure 
is computationally expensive. 
More realistically, network constraints may limit the maximum chatting rate,
which greatly reduces the search space.

\begin{figure}
	\centering
	\begin{tabular}{cc}
	\includegraphics[width=2.2in]{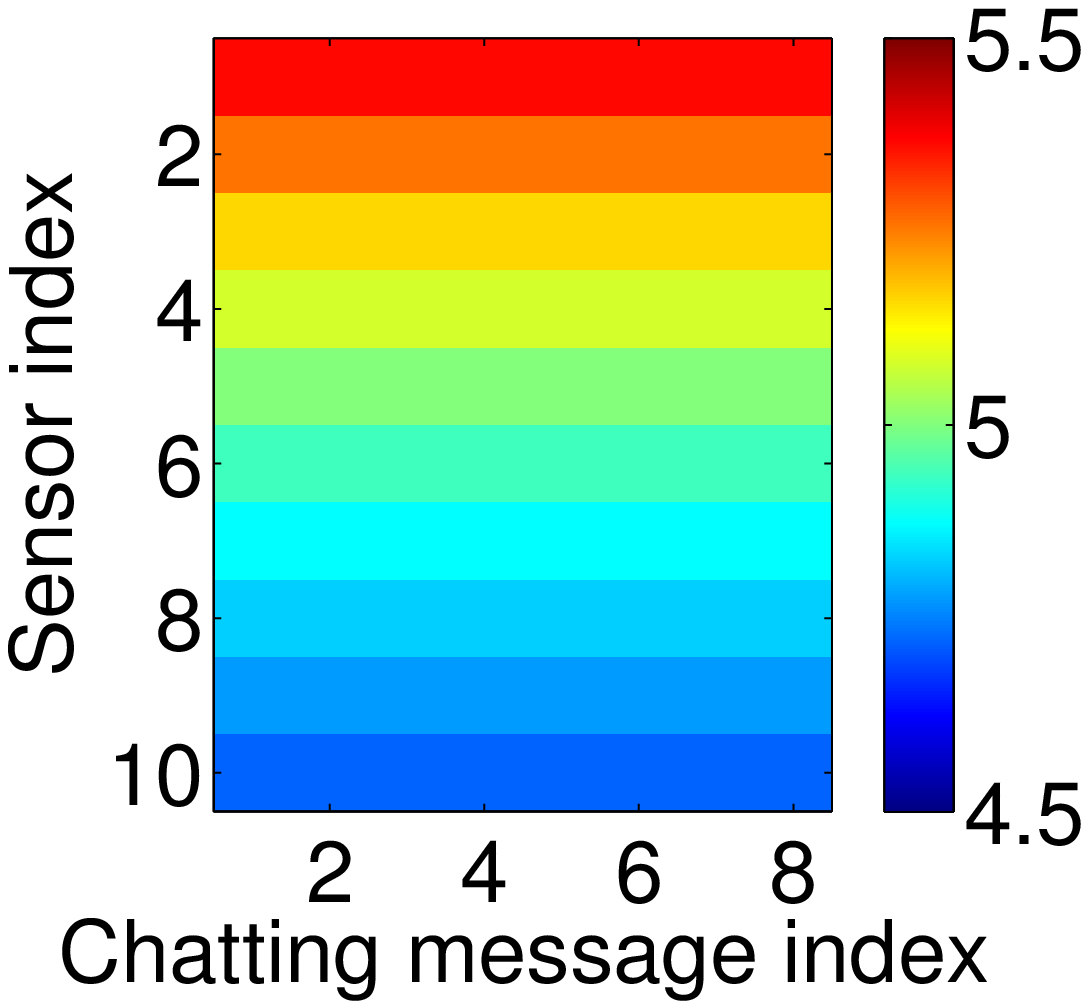} & 
				\includegraphics[width=2.2in]{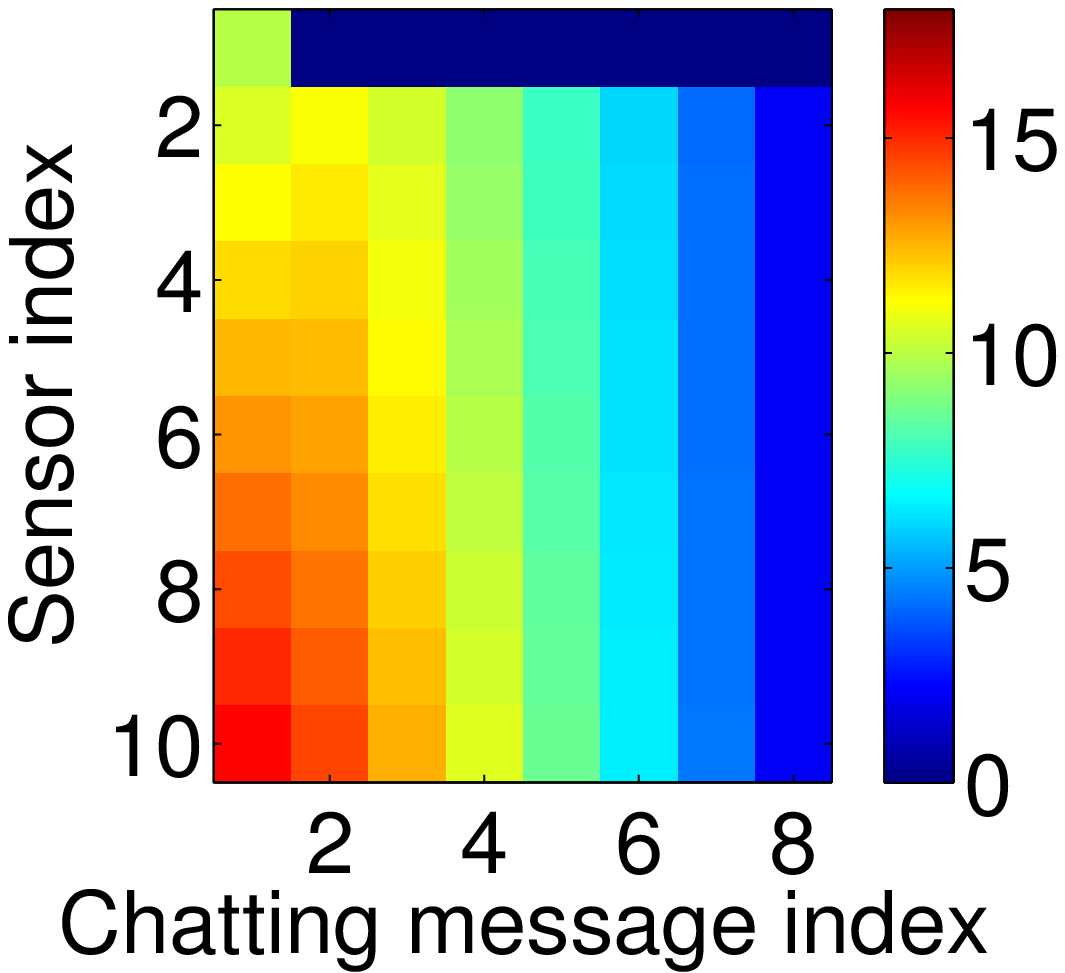}\\
	{\small (a)} & {\small (b)} 
	\end{tabular}
	\caption{ 
	  	Cost allocation for a maximum computation network, as described in Section~\ref{sec:max}. 
	  	In this case, $N = 10$, $C = 5 N$, $R_c = 3$, $\alpha_c = 0$, and $\alpha_n = 1$.
	  	In the fixed-rate setting (a), the sensors are allowed to have different communication rates but cannot 
	  	adjust the rate with the received chatting message.
	  	In the entropy-constrained setting (b), each sensor except sensor 1 receives chatting messages and 
	  	can adjust its communication rate appropriately.  }
	\label{fig:allocation}
\end{figure}

In Fig.~\ref{fig:allocation}, we show optimal communication rates for the network described in
Section~\ref{sec:max}.  
We delay description of the specific network properties and aim only to illustrate how the 
cost allocations $b_n(m)$ may change depending with sensors or chatting messages. 
Under fixed-rate coding, $b_n$ varies depending on the chatting graph.
In the entropy-constrained setting, the allocation can also vary with the chatting messages, 
except for Sensor~1\@.
This increased flexibility allows for a wider range of rates, as well as improved performance in many situations.

\section{Maximum Computation}
\label{sec:max}

The results in the previous sections hold generally, and we now build some intuition about chatting using
a specific distributed network performing a maximum computation. 
The choice of this computation is not arbitrary; we will show that it allows for a particular chatting 
architecture that makes it convenient to study large networks. 
Moreover, this network reveals some surprising insights into the behavior of chatting. 
While this paper restricts its attention solely to the maximum computation, 
more examples are discussed in~\cite{SunG:12}.

\subsection{Problem Model}
\label{sec:max:model}

\begin{figure}
  \centering
	\includegraphics[width=4in]{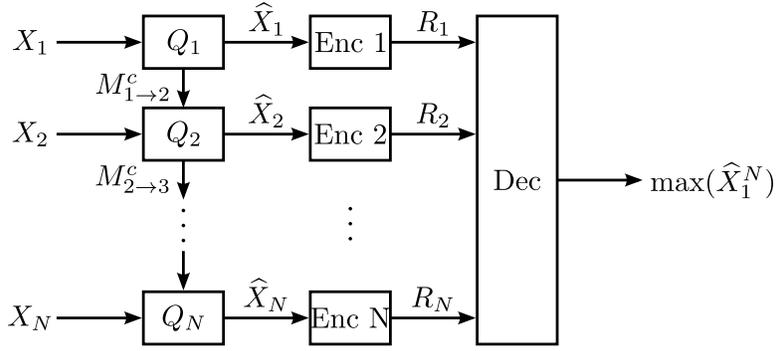}
  \caption{A fusion center wishes to determine the maximum of $N$ iid uniform sources and receives messages
  						$M_n$ from each sensor $n$ at rate $R_n$.
  						The sensors are allowed to chat serially down the network
  						using messages $M_{(n-1)\to n}$ at rate $R_c$. }
  \label{fig:maxex}
\end{figure}

We consider a network where the fusion center aims to reproduce the maximum of $N$ sources, 
where each $X_n$ is independent and uniformly distributed on $[0,1]$.
The sensors measuring these sources are allowed to chat in a serial chain, meaning each sensor has at most 
one parent and one child (see Fig.~\ref{fig:maxex}). 
Initially, we will consider the simplest such network with the following assumptions:
\begin{enumerate}
	\item The chatting is serial, meaning the sequence of chatting messages is $\{ M_{(n-1) \to n} \}_{n=2}^N$.
	\item Each chatting link is identical and has rate $R_c$, codebook size $K_c = 2^{R_c}$ and cost $\alpha_c$.
	\item The communication links between sensors and the fusion center are allowed to have different rates. 
					For simplicity, we assume them to be homogeneous and normalize the cost to be $\alpha_n = 1$.
	\item The outgoing chatting message at Sensor~$1$ is the index of a uniformly quantized version of 
						its observation with $K_c$ levels.
	\item For $n > 1$, the chatting message from Sensor~$n$ is the maximum of the index of 
							Sensor~$n$'s own uniformly quantized observation and the chatting message from its parent.
\end{enumerate}

Under this architecture, the chatting messages effectively correspond to a uniformly quantized observation
of the maximum of all ancestor nodes:
\beq 
	\label{eq:max:chatmessages}
	M_{(n-1) \to n} = \mathcal{I}(Q_{K_c, U} (\max(X_1^{n-1}))) , 
\eeq
where $\mathcal{I}$ is the index of the quantization codeword and can takes values $\{1 , \ldots , K_c\}$.
The simplicity of the chatting message here arises from the permutation-invariance of the maximum function.
We will exploit this structure to provide precise characterizations of system performance. 

\subsection{Quantizer Design}
\label{sec:max:design}

Using~\eqref{eq:gamman}, we find the max function has sensitivity $\gamma_n^2(x) = x^{N-1}$ for all $n$.
Without chatting, each sensor's quantizer would be the same with a point density that is a function of the 
source distribution and sensitivity. 
Moreover, since the cost per bit of transmitting to the fusion center is the same, the solution of the resource
allocation problem assigns equal weight to each link. 
Hence, minimizing~\eqref{eq:dfsq:distfr} yields the optimal fixed-rate distortion--cost trade-off:
\beq
	\label{eq:max:nochatFRD}
	D_\mathrm{max,fr} (C) \APPROX \frac{N}{12} \left( \frac{3}{N+2} \right)^3 2^{C/N} . 
\eeq
Similarly, the minimum of~\eqref{eq:dfsq:distec} leads to the optimal entropy-constrained 
distortion--cost trade-off
\beq
	\label{eq:max:nochatECD}
	D_\mathrm{max,ec} (C) \APPROX \frac{N}{12} e^{-N+1} 2^{C/N} .
\eeq
These high-resolution expressions provide scaling laws on how the distortion relates to the number of sensors. They require the total cost $C$ increase linearly with $N$ to hold. 

With chatting, we first need to determine the conditional sensitivity, which is given below for uniform sources:
\begin{lemma}
	\label{lem:maxex}
	Given $K_c = 2^{R_c}$, the sensitivity profile corresponding to a received chatting message 
	$M_{(n-1) \to n} = k$ is
	\beq
		\gamma_{n \, | \, M_{(n-1) \to n}}^2 (x \, | \, k) = \left\{
     	\begin{array}{ll}
      	 0 , &  x  < \frac{k-1}{K_c} ; \\
       		\frac{(K_c x)^{n-1} - (k-1)^{n-1}}{k^{n-1} - (k-1)^{n-1}} x^{N-n} , 
       																	& \frac{k-1}{K_c} \leq x < \frac{k}{K_c} ; \\
       		x^{N-n} , & x \geq \frac{k}{K_c} .
     \end{array}
   \right.
  \eeq
\end{lemma}	
\begin{IEEEproof}		
	See Appendix~\ref{app:cor:unifmax}.
\end{IEEEproof}

We have already noted the incident chatting message of Sensor~$n$ is a uniformly quantized observation of
$Y_n = \max(X_1^{n-1})$, where $f_Y(y) = (n-1) y^{n-2}$. 
Hence, 
\beq
	\label{eq:prob_message_max}
	\Prb \left( M_{(n-1) \to n} = k \right) = \left(\frac{k}{K_c} \right)^{n-1} - \left(\frac{k-1}{K_c} \right)^{n-1} .
\eeq
Below, we give distortion asymptotics for the serial chatting network under both fixed-rate and
entropy-constrained quantization.

\subsubsection{Fixed-rate case}

From Theorem~\ref{thm:chatdist}, the asymptotic total fMSE distortion is
\beq
	\label{eq:totaldistFR}
	\sum_{n=1}^N \beta_n 2^{-2 R_n} ,
\eeq
where $\beta_n = \frac{1}{12} \| \gamma^2_{n|M^c} \|_{1/3}$.
Because Sensor~1 has no incoming chatting messages, its sensitivity is $\gamma_1^2(x) = x^{N-1}$ and
the resulting distortion constant is
\[ \beta_1 = \frac{1}{12} \left( \frac{3}{N+2} \right)^3 . \]
For other sensors, the distortion contribution is
\[ \beta_n = \frac{1}{12} \sum_{k=1}^{K_c}
			\Prb \left( M_{(n-1) \to n} = k \right) \, \big\| \gamma_{n \, | \, M_{(n-1) \to n} = k}^2 \big\|_{1/3} .
\]
For Sensor~$n$ with $n > 1$, all incoming messages besides $k = 1$ induce a don't-care interval, 
so one of the $2^{R_n}$ codewords is placed exactly at $\Frac{(k-1)}{K}$.

We study the trade-off between chatting rate $R_c$ and fMSE for several choices of $N$ and $\alpha_c$ 
using optimal cost allocation as determined by Lemma~\ref{lem:costop:lagrange}.
In Fig.~\ref{fig:chattingfMSE}a, we observe that increasing the chatting rate yields improvements in fMSE\@.  
As the number of sensors increases, this improvement becomes more pronounced.  
However, this is contingent on the chatting cost $\alpha_c$ being low.  
As discussed in Section~\ref{sec:prelim:dfsq}, chatting can lead to worse system performance if
the cost of chatting is on the same order as the cost of communication given a total resource budget,
as demonstrated by Fig.~\ref{fig:chattingfMSE}c.
Although the main results of this work are asymptotic, we have asserted the distortion equations are 
reasonable at finite rates. 
To demonstrate this, we design real quantizers under the same cost constraint and demonstrate that the resulting
performance is comparable to high-resolution approximations of Theorem~\ref{thm:chatdist}.
This is observed in Figs.~\ref{fig:chattingfMSE}a and c, which shows the asymptotic prediction of the 
distortion--rate trade-off is accurate even at 4 bits/sample.

\subsubsection{Entropy-constrained case}

Generally, the total distortion in the entropy-constrained case is
\beq
	\sum_{n=1}^N \E \left[ \beta_{n,k} 2^{-2 R_{n,k}} \, \middle| \, M_{(n-1) \to n} = k \right] ,
\eeq
noting each sensor is allowed to vary its communication rate with the chatting messages it receives.
Like in the fixed-rate setting, an incoming message $k$ will induce a don't-care interval of
$[0,\Frac{(k-1)}{K}]$ in the conditional sensitivity.
If $A_{n,k}$ is the event that $X_n$ is not in a don't-care interval when receiving message $k$, then
\beq
	\beta_{n,k} = \frac{1}{12} \Prb \left( M_{(n-1) \to n} = k \right) \,
								2^{2 h(X_n|A_{n,k}) + 2 \E [\log_2 \gamma_{n \, | \, M_{(n-1) \to n}} (X_n|k)]}
\eeq
and $R_{n,k} = \Frac{(R_n- H_B(\Prb(A_{n,k})))}{\Prb(A_{n,k})}$.

\begin{figure*}
	\centering
	\begin{tabular}{cc}
	\includegraphics[width=2.5in]{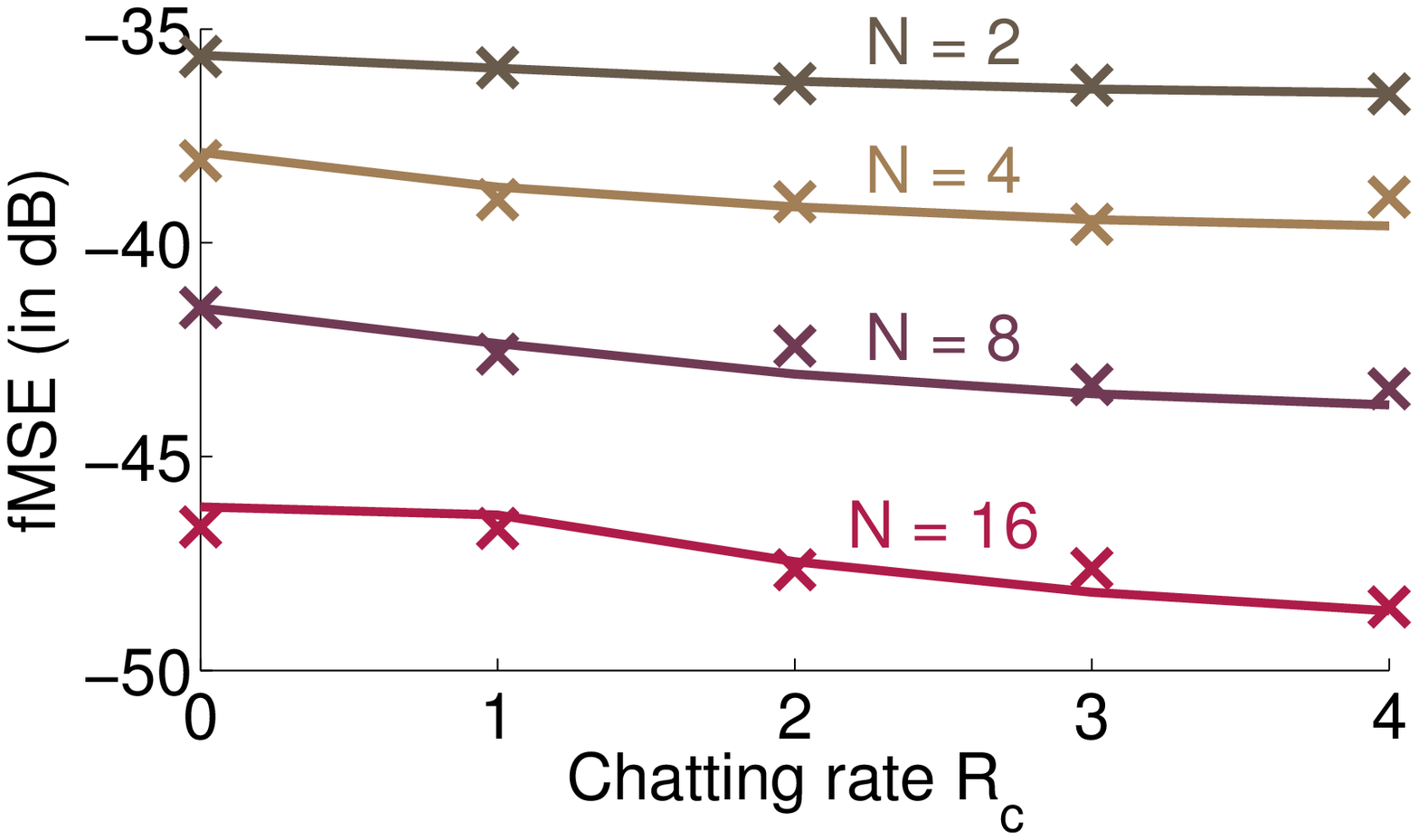} & 
				\includegraphics[width=2.5in]{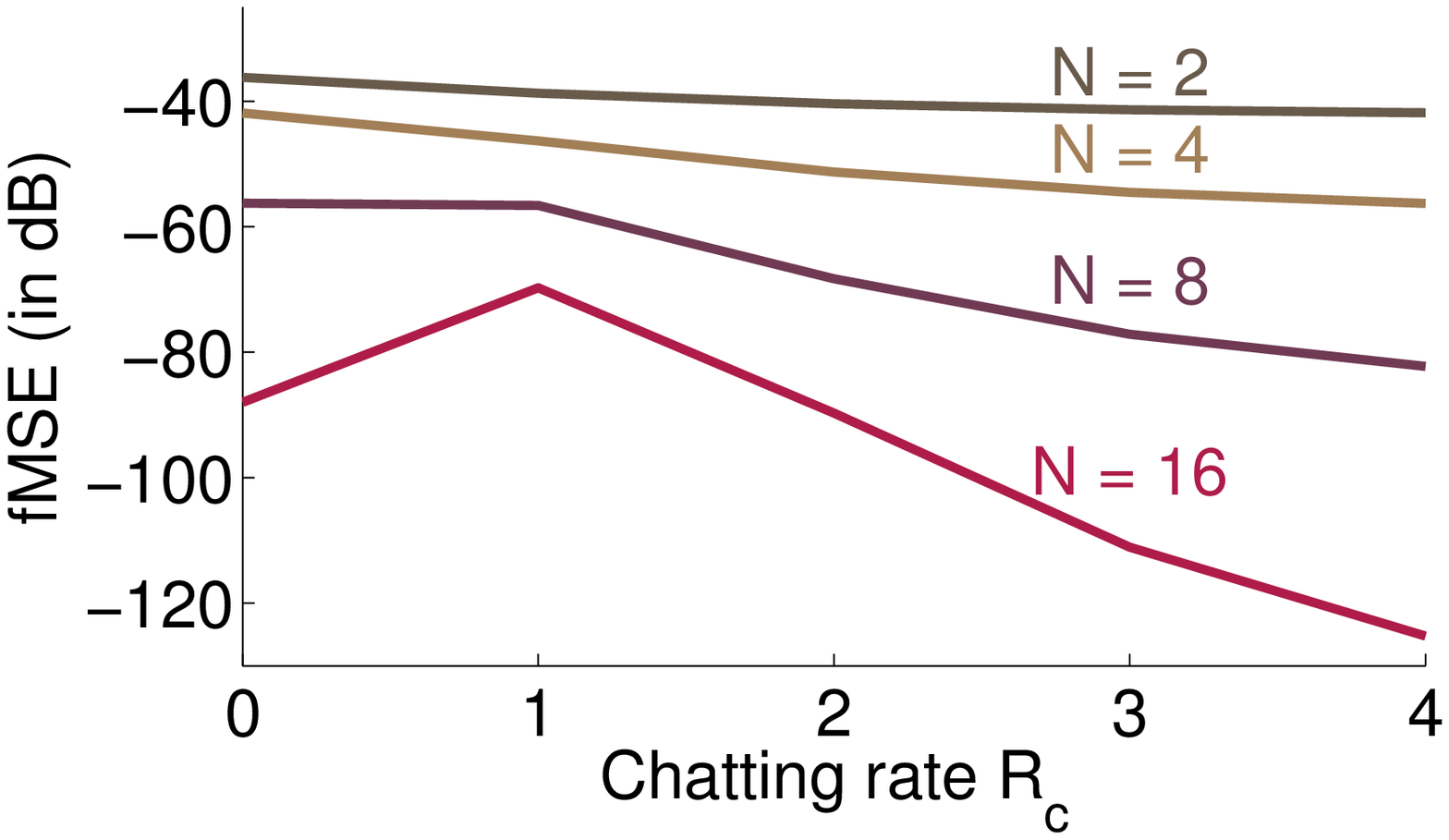}\\
	{\small (a)} & {\small (b)} \\
	\includegraphics[width=2.5in]{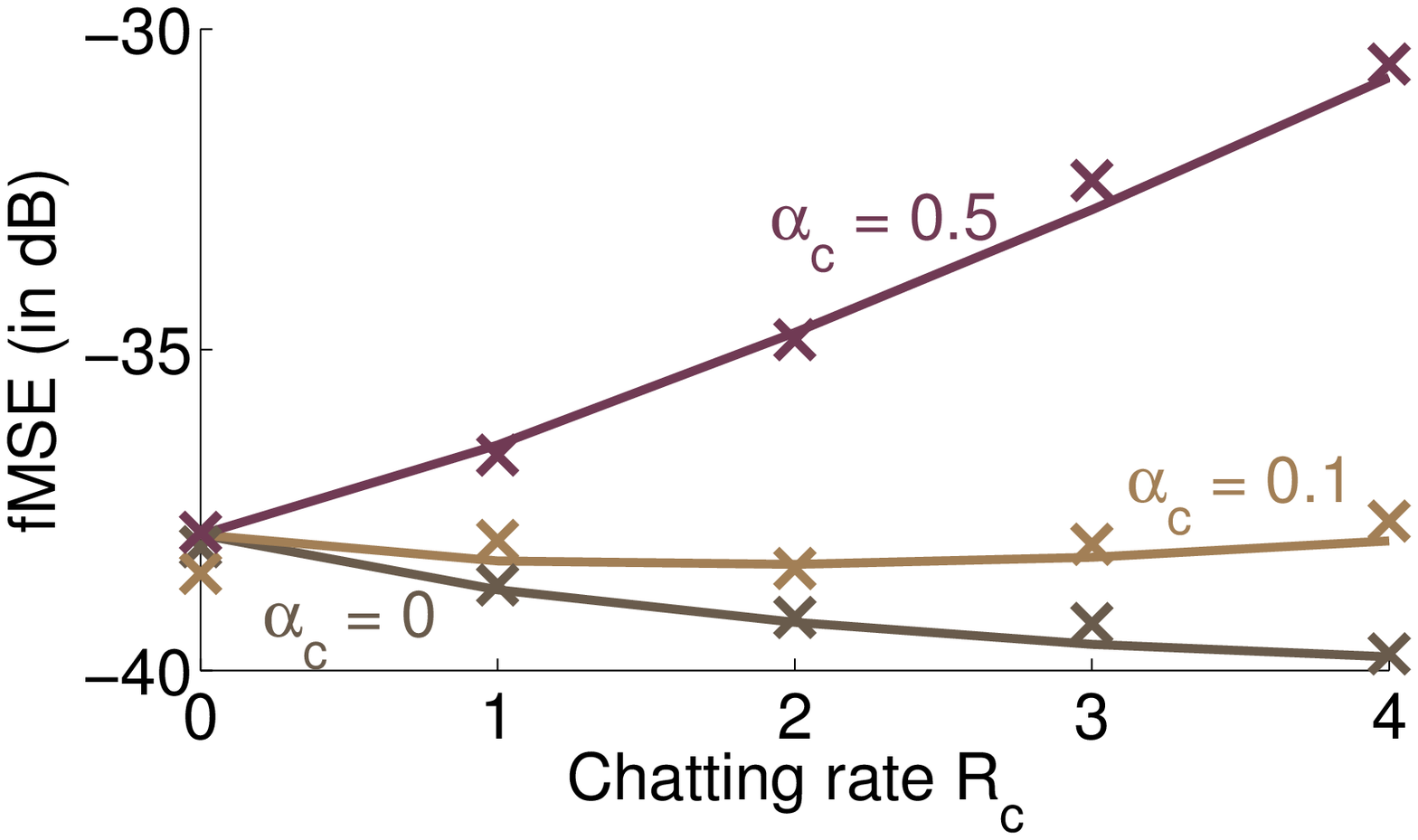} & 
				\includegraphics[width=2.5in]{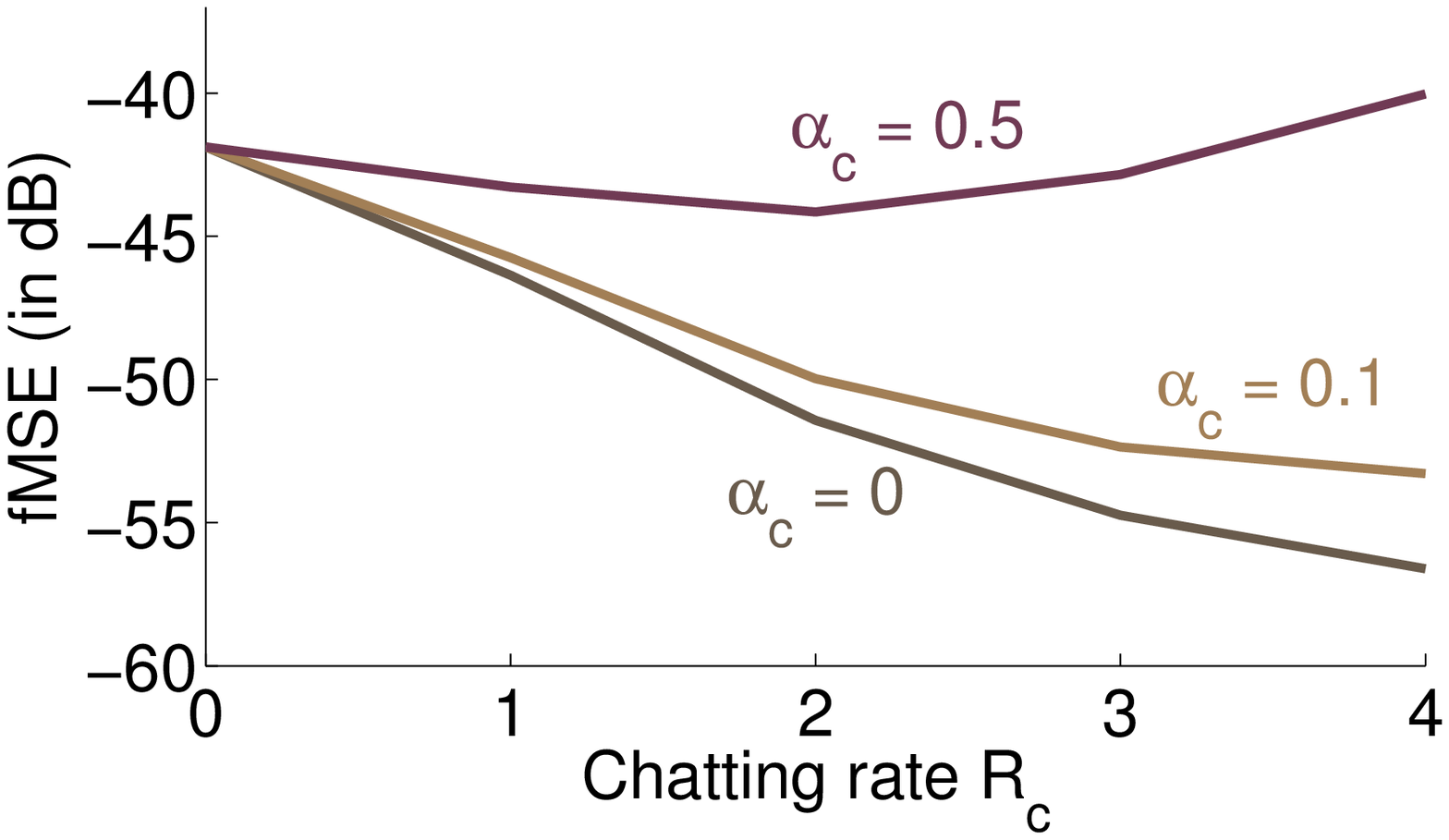} \\
	{\small (c)} & {\small (d)}
	\end{tabular}
	\caption{ 
	  	Performance of the maximum computation network in both the fixed-rate (left plots)
	  	and entropy-constrained (right plots) settings.
	  	Plots (a) and (b) illustrate the trade-off between fMSE and chatting rate for choices of $N$
	  	assuming total cost $C = 4 N$ and $\alpha_c = 0.01$.
	  	Plots (c) and (d) illustrate the trade-off between fMSE and chatting rate for choices of $\alpha_c$
	  	assuming $N = 4$ sensors and total cost $C = 4N$.
	  	In all cases, the cost of communication is $\alpha_n = 1$.
	  	For the fixed-rate setting, we validate the distortion through simulated runs on real quantizers
	  	designed using~\eqref{eq:chat:lamfr}.
	  	We observe that high-resolution theory predicts actual performance at rates as low as 4 bits/sample,
	  	as shown by crosses in the fixed-rate plots.
	  	 }
	\label{fig:chattingfMSE}
\end{figure*}

Like in the fixed-rate setting, we study the relationship between the chatting rate $R_c$ and fMSE,
this time using the probabilistic allocation optimization of Lemma~\ref{lem:costop:lagrange_prob} 
in Appendix~\ref{app:allocation}.
Due to the extra flexibility of allowing a sensor to vary its communication to the fusion center 
with the chatting messages it receives,
we observe that increasing the chatting rate can improve performance more dramatically than in the 
fixed-rate case (see Fig.~\ref{fig:chattingfMSE}b).
Surprisingly, chatting can also lead to inferior performance for some combinations of $R_c$ and $N$,
even when $\alpha_c$ is small.  
This phenomenon will be discussed in greater detail below.
In Fig.~\ref{fig:chattingfMSE}d, 
we compare different choices of $\alpha_c$ to see how performance changes with
the chatting rate.  
Unlike for fixed rate, in the entropy-constrained setting, chatting can be useful even when
its cost is close to the cost of communication to the fusion center. 

\subsection{Generalizing the Chatting Messages}
\label{sec:max:chatting}

We have considered the case where a chatting message
is the uniform quantization of the maximum of all ancestor nodes, as shown in~\eqref{eq:max:chatmessages}. 
Although simple, this coding of chatting messages is not optimal.  
Here, we generalize chatting messages to understand how the performance can change 
with this design choice.

\begin{figure*}
	\centering
	\begin{tabular}{cc}
	\includegraphics[width=2.5in]{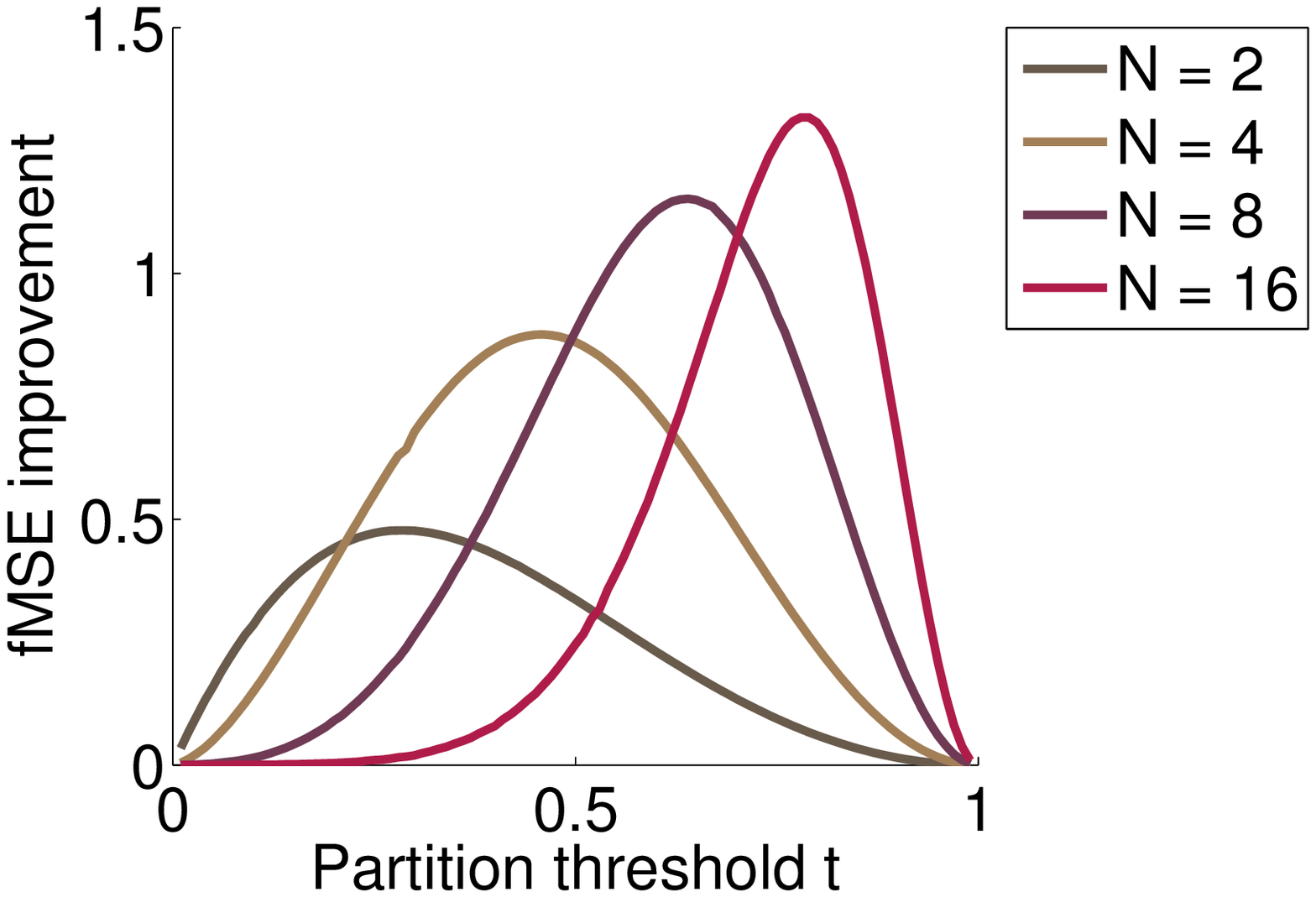} & 
				\includegraphics[width=2.5in]{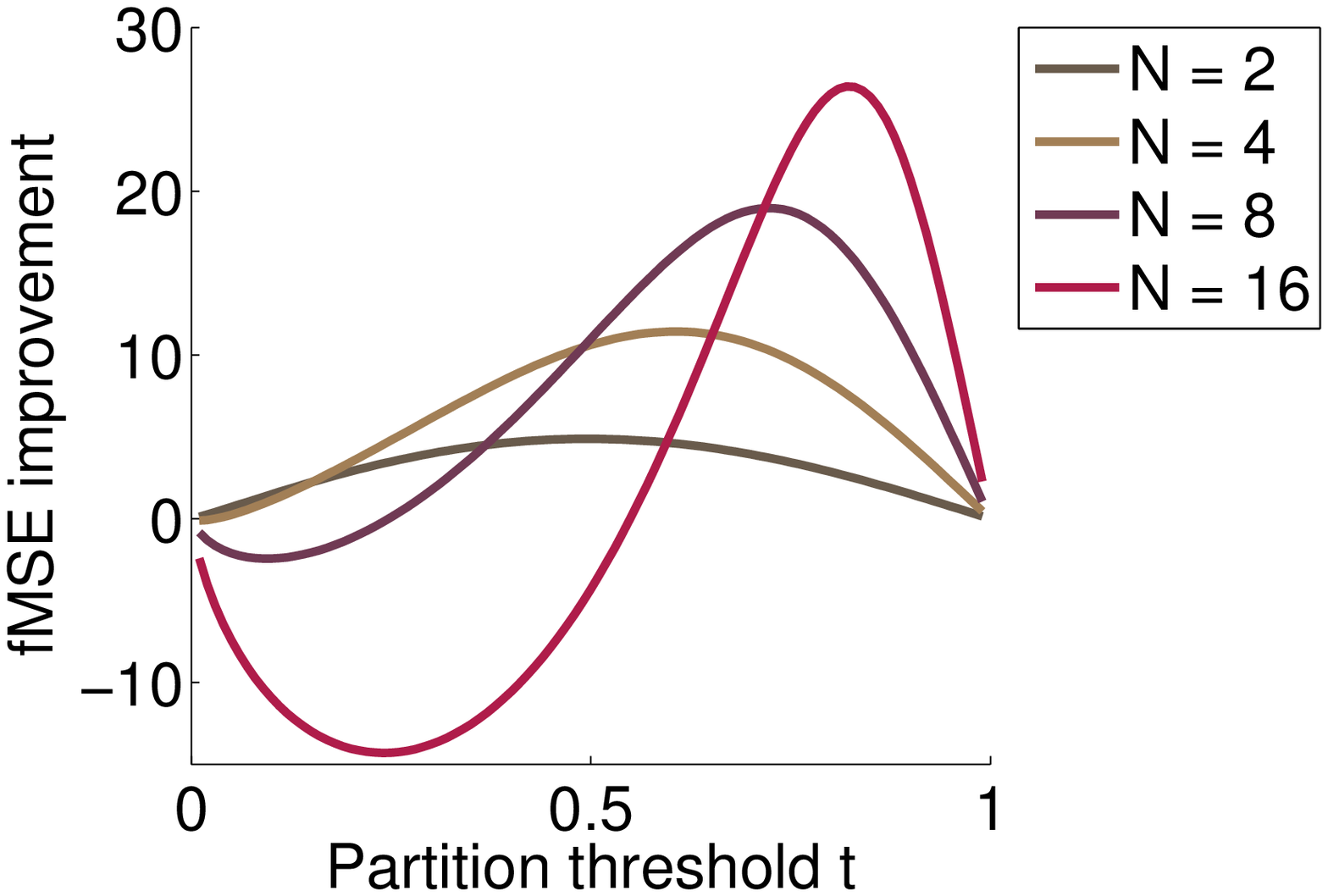}\\
	{\small (a)} & {\small (b)} \\
	\end{tabular}
	\caption{ 
	  	Distortion improvement compared to no chatting in the maximum computation network for
	  	the fixed-rate (left plot) and entropy-constrained (right plot) settings when 
	  	varying the partition boundary $p_1$. 
	  	We assume chatting is free, i.e., $\alpha_c = 0$, but the chatting 
	  	rate is limited to one bit. }
	\label{fig:genchat}
\end{figure*}

We begin by considering the same network under the restriction that the chatting rate is $R_c = 1$, 
but allow the single partition boundary $p_1$ to vary rather than setting it to $1/2$. 
Currently, we keep the coding consistent for every sensor such that a chatting message $k=1$ implies
$\max(X_1^{n-1}) \in [0,p_1]$ and $k=2$ means $\max(X_1^{n-1}) \in (p_1,1]$.
Distortions for a range of $N$ and $p_1$ are shown in Fig.~\ref{fig:genchat}.

From these performance results, we see that the choice of $p_1$ should increase with the size of the network,
but precise characterization of the best $p_1$ is difficult because of the complicated effect the 
conditional sensitivity has on both the distortion constants and rate allocation.
We can recover some of the results of Fig.~\ref{fig:chattingfMSE} by considering $p_1=1/2$.
It is now evident that this choice of $p_1$ can be very suboptimal, especially as $N$ becomes large.
In fact, we observe that for certain choices of the partition with entropy coding, the distortion with 
chatting can be larger than from a traditional distributed network even though the chatting cost is 0.
This unintuitive fact arises because the system's reliance on the conditional sensitivity is fixed,
and the benefits of a don't-care interval are mitigated by creating a more unfavorable conditional sensitivity.
We emphasize that this phenomenon disappears as the rate becomes very large.

Since the flexibility in the choice of the chatting encoder's partitions can lead to improved performance 
when $R_c = 1$, we can expect even more gains when the chatting rate is increased.
However, the only method for optimizing the choice of partition boundaries developed currently 
involve brute-force search using the conditional sensitivity derived in Appendix~\ref{app:cor:unifmax}.
Another extension that leads to improved performance is to allow chatting encoders to employ 
different partitions.
This more general framework yields strictly improved results, but some of the special structure of the
serial chatting network is lost as the chatting message is no longer necessarily the maximum of all 
ancestor sensors. 
The added complexity of either of these extensions make their performances difficult to quantify. 

\subsection{Optimizing a Chatting Network}
\label{sec:max:optimize}

In this paper, we have formulated a framework allowing low-rate collaboration between sensors in a 
distributed network.
We have introduced several methods to optimize such a network, including 
nonuniform quantization, rate allocation, and design of chatting messages.
Here, we combine these ingredients and see how each one impacts fMSE\@.

We will continue working with the maximum computation network from Fig.~\ref{fig:maxex}
assuming $R_c = 1$, $\alpha_c = 0$, $N = 5$ and $C = 5 N$. 
We further assume the coding of chatting messages is the same for every sensor on the serial chain. 
We will then consider the following scenarios:
\begin{enumerate}
	\item A chatting network with $R_n = 5$ for all $n$ and chatting designed by~\eqref{eq:max:chatmessages}.
	\item A chatting network with rate allocation and chatting designed by~\eqref{eq:max:chatmessages}.
	\item A chatting network with rate allocation and optimization over chatting messages.
\end{enumerate}

\begin{figure}
	\centering
	\includegraphics[width=2.5in]{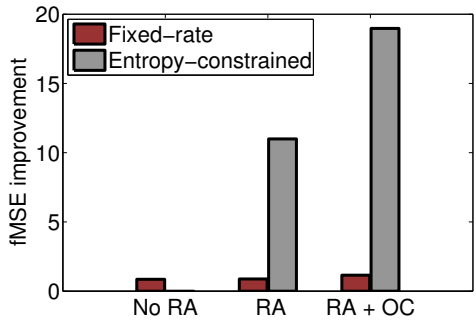}
	\caption{ 
	  	Distortion improvement for Scenarios 1--3 over a distributed network without chatting.  
	  	Both rate allocation (RA) and optimized chatting (OC) are considered. }
	\label{fig:overallopp}
\end{figure}

We analyze the fMSE of each scenario compared to a distributed network without chatting ($R_c = 0$).
From Fig.~\ref{fig:overallopp}, we can see that incorporating rate allocation and chatting optimization
yields substantial gains in the entropy-constrained setting.  
For fixed rate, the most meaningful improvement comes from allowing chatting, while
additional optimization provides little additional benefit.
Up to this point, we have limited chatting to have fixed codebook size and did not allow entropy coding.
Lifting these restrictions increase system complexity and can provide even greater compression gain. 

\section{Conclusions}
\label{sec:conclude}

In this work, we explored how intersensor communication---termed \emph{chatting}---can improve 
approximation of a function of sensed data in a distributed network constrained to scalar quantization. 
We have motivated chatting from two directions: 
providing an analysis technique for distortion performance when low-blocklength limitations make
Shannon theory too optimistic, 
and illustrating the potential gains over simplistic practical designs.
There are many opportunities to leverage heterogeneous network design to aid information acquisition
using the tools of high-resolution theory,
and we provide precise characterizations of distortion performance, quantizer design, and cost allocation
to optimize distributed networks. 
Many challenges remain in analyzing chatting networks. 
Some future directions that are meaningful include a more systematic understanding of how to design 
chatting messages and applications where chatting may be feasible and beneficial. 

One can consider ``sensors'' being distributed in time rather than space, with the decoder computing a function of samples from a random process. 
Connections of this formulation to structured vector quantizers are of independent interest.

\appendices

\section{Rate Allocation for Distributed Networks}
\label{app:allocation}

Consider the distributed network in Fig.~\ref{fig:star} without the chatting channel.
The cost per bit of the communication link and the cost allocation between Sensor~$n$ and the fusion center is
denoted by $\alpha_n$ and $b_n$ respectively, leading to a communication rate of $R_n = b_n/\alpha_n$.
Below, we solve the cost allocation problem under the assumption that companding quantizers are used 
and noninteger rates are allowed.

\begin{lemma}
	\label{lem:costop:kkt}
	The optimal solution to 
	\beq
		\label{eq:distcostop}
		D(C) = \min_{\sum b_n = C, b_n \geq 0} \sum_{n=1}^N \beta_n 2^{-2 b_n / \alpha_n}
	\eeq
	has cost allocation
	\beq
		\label{eq:distcostop:b}
		b_n^* = \max \left( 0, \frac{1}{2} \log_2 \frac{\beta_n/\alpha_n}{\tilde{\beta}} \right) , 
	\eeq
	where $\tilde{\beta}$ is chosen such that $\sum b_n^* = C$. 
\end{lemma}
\begin{IEEEproof}
	This lemma extends the result from~\cite{Segall:76} or can be derived directly from the KKT conditions.
\end{IEEEproof}

Each $\beta_n$ is calculated using only the functional sensitivity $\gamma_n$ and marginal source
pdf $f_{X_n}$.
Although Lemma~\ref{lem:costop:kkt} is always true, we emphasize that its effectiveness in predicting the 
proper cost allocation in a distributed network is only rigorously shown for high cost (i.e. high rate) 
due to its dependence on~\eqref{eq:fmse:dist}.
However, it can be experimentally verified that costs corresponding to moderate communication rates still 
yield near-optimal allocations. 

When the solution of Lemma~\ref{lem:costop:kkt} is strictly positive, a closed-form expression exists:
\begin{lemma}
	\label{lem:costop:lagrange}
	Assuming each $b_n^*$ in~\eqref{eq:distcostop:b} is strictly positive, it can be expressed as
	\beq
		b_n^* = \frac{\alpha_n}{\tilde{\alpha}} C 
							+ \frac{\alpha_n}{2} \log_2 \frac{\beta_n/\alpha_n}
											{\left( \prod_j \left(\beta_j/\alpha_j \right)^{\alpha_j} \right)^{1/\sum \alpha_i} } .
	\eeq
\end{lemma}

\begin{IEEEproof}
	The proof uses Lagrangian optimization.
\end{IEEEproof}

If Sensor~$n$ is allowed to vary the communication rate depending on the side information $M_{\mathrm{si},n}$ 
it receives, further gains can be enjoyed.
This situation is natural in chatting networks, where the side information is the low-rate messages passed
by neighboring sensors. 
Here, we introduce \emph{probabilistic cost allocation}, yielding a distortion--cost trade-off
\beq
	\label{eq:distcostop_prob}
	D(C) = \min_{\substack{\sum \E[b_n(M_{\mathrm{si},n})] = C \\ b_n(m) \geq 0}}
						\sum_{n=1}^N \E\left[ \beta_n(M_{\mathrm{si},n}) 2^{-2 b_n(M_{\mathrm{si},n}) / \alpha_n} \right] ,
\eeq
where the expectation is taken with respect to $M_{\mathrm{si},n}$.
Each link will have a cost allocation $b_n(m)$ for every possible message $m$ while satisfying 
an average cost constraint.
An analogous result to Lemma~\ref{lem:costop:kkt} can be derived;
for the situation where the optimal allocation is strictly positive, it can again be expressed in closed form:
\begin{lemma}
	\label{lem:costop:lagrange_prob}
	Assume the side information $M_{\mathrm{si},n}$ received at Sensor~$n$ is $m \in \mathcal{M}_n$
	and the cost per bit of the communication link may vary with $m$.
	Assuming each allocation $b^*_n(m)$ in the solution to~\eqref{eq:distcostop_prob} is strictly positive, 
	it can be expressed as
	\beq
		b_n^*(m) = \frac{\alpha_n(m)}{\tilde{\alpha}} C 
							+ \frac{\alpha_n(m)}{2} \log_2 \frac{\beta_n(m)/\alpha_n(m)}
											{\prod_j \prod_l \left( \left(\beta_j(l)/\alpha_j(l) \right)^{\alpha_j(l) /\tilde{\alpha}} \right) } ,
	\eeq
	where $\tilde{\alpha} = \sum_n \sum_m f_{M_{\mathrm{si},n}}(m) \, \alpha_n(m)$.
\end{lemma}

Here, we extended previous known rate allocation results~\cite{Segall:76,GershoG:92} to 
account for heterogeneity in distributed networks. 
Although these results do not account for chatting, we see in Section~\ref{sec:chatall} that they become important 
tools in optimizing performance in such networks.

\section{Sensitivity of Maximum Computation Network}
\label{app:cor:unifmax}

Assuming iid uniform sources on the support $[0,1]$, the sensitivity of each sensor in the 
maximum computation network in Fig.~\ref{fig:maxex} without chatting is
\begin{align*}
	\gamma_n^2(x) &= \E [ |g_n(X_1^N)|^2 \, | \, X_n = x] \\
		&= \Prb \left( \min(X_1^N) = X_n \, | \, X_n = x \right) \\
		&= \Prb(X_1 < x) \cdots \Prb(X_{n-1} < x) \Prb(X_{n+1} < x) \cdots \Prb(X_N < x) \\
		&= x^{N-1} .
\end{align*}

When the chatting graph is a serial chain, Sensor~$n$ has some lossy version of the information collected 
by its ancestor sensors.
For the max function, chatting reduces the support of the estimate of $\max(X_1^{n-1})$ by Sensor~$n$. 
Hence, the message $M_{(n-1)\to n}$ reveals the max of the ancestor sensors is
in the range $[s_l, s_u]$. 
This side information forms three distinct intervals in the conditional sensitivity. 
First, in the interval $x < s_l$, $X_n$ is assuredly less than $\max(X_1^{n-1})$ and hence
sensitivity is 0 since the information at Sensor~$n$ is irrelevant at the fusion center.
Second, if $x > s_u$, $X_n$ is greater than $\max(X_1^{n-1})$ and the sensitivity 
should only depend on the number of descendant sensors, 
leading to a sensitivity of $x^{N-n}$. 
Finally, when $s_l \leq x < s_u$,
Sensor~$n$ must take into consideration both ancestors and descendants, yielding sensitivity
\begin{align*}
	& \Prb \left( \min(X_1^N) = X_n \, \middle| \, X_n = x, \max(X_1^{n-1}) \in [s_l, s_u] \right) \\
		& \hspace{12ex} = \Prb\left( \max(X_1^{n-1}) < x \, \middle| \, \max(X_1^{n-1}) \in [s_l, s_u] \right) \,
																		\Prb \left( \max(X_{n+1}^N) < x \right) \\
		& \hspace{12ex} =  \frac{x^{n-1} - s_l^{n-1}}{s_u^{n-1} - s_l^{n-1}} x^{N-n} .
\end{align*}
More specific to the case when messages correspond to uniform quantization, we define
$K_c = 2^{R_c}$ and denote each received message $M_{(n-1)\to n}$ as $k_n$. 
Setting $s_l = \Frac{(k_n-1)}{K_c}$ and $s_u = \Frac{k_n}{K_c}$ gives Lemma~\ref{lem:maxex}.

\bibliographystyle{ieeetr}
\bibliography{bib_dsc,bib_quant,bib_comm}

\begin{thebibliography}{10}

\bibitem{SlepianW:73}
D.~Slepian and J.~K. Wolf, ``Noiseless coding of correlated information
  sources,'' {\em IEEE Trans. Inform. Theory}, vol.~IT-19, pp.~471--480, July
  1973.

\bibitem{Zamir:96}
R.~Zamir, ``The rate loss in the {W}yner--{Z}iv problem,'' {\em IEEE Trans.
  Inform. Theory}, vol.~42, pp.~2073--2084, Nov. 1996.

\bibitem{TanK:12arxiv}
V.~Y.~F. Tan and O.~Kosut, ``On the dispersions of three network information
  theory problems.'' arXiv:1201.3901v2 [cs.IT]., Feb. 2012.

\bibitem{MisraGV:11}
V.~Misra, V.~K. Goyal, and L.~R. Varshney, ``Distributed scalar quantization
  for computing: High-resolution analysis and extensions,'' {\em IEEE Trans.
  Inform. Theory}, vol.~57, pp.~5298--5325, Aug. 2011.

\bibitem{SunMG:12arxiv}
J.~Z. Sun, V.~Misra, and V.~K. Goyal, ``Distributed functional scalar
  quantization simplified.'' arXiv:1206.1299v1 [cs.IT]., June 2012.

\bibitem{TseV:05}
D.~Tse and P.~Viswanath, {\em Fundamentals of Wireless Communication}.
\newblock Cambridge, UK: Cambridge University Press, 2005.

\bibitem{YucekA:09}
T.~Yucek and H.~Arslan, ``A survey of spectrum sensing algorithms for cognitive
  radio applications,'' {\em IEEE Comm. Surveys Tutorials}, vol.~11, no.~1,
  pp.~116--130, 2009.

\bibitem{SunG:12}
J.~Z. Sun and V.~K. Goyal, ``Chatting in distributed quantization networks,''
  in {\em Proc. 50th Ann. Allerton Conf. on Commun., Control and Comp.},
  (Monticello, IL), Oct. 2012.

\bibitem{OrlitskyR:01}
A.~Orlitsky and J.~R. Roche, ``Coding for computing,'' {\em IEEE Trans. Inform.
  Theory}, vol.~47, pp.~903--917, Mar. 2001.

\bibitem{FengES:04}
H.~Feng, M.~Effros, and S.~A. Savari, ``Functional source coding for networks
  with receiver side information,'' in {\em Proc. 42nd Annu. Allerton Conf.
  Commun. Control Comput.}, pp.~1419--1427, Sept. 2004.

\bibitem{PolyanskiyPV:10}
Y.~Polyanskiy, H.~V. Poor, and S.~Verdu, ``Channel coding rate in the finite
  blocklength regime,'' {\em IEEE Trans. Inform. Theory}, vol.~56,
  pp.~2307--2359, May 2010.

\bibitem{IngberK:11}
A.~Ingber and Y.~Kochman, ``The dispersion of lossy source coding,'' in {\em
  Proc. IEEE Data Compression Conf.}, (Snowbird, Utah), pp.~53--62, Mar. 2011.

\bibitem{KostinaV:12}
V.~Kostina and S.~Verdu, ``Fixed-length lossy compression in the finite
  blocklength regime,'' {\em IEEE Trans. Inform. Theory}, vol.~58,
  pp.~3309--3338, June 2012.

\bibitem{KaspiB:82}
A.~H. Kaspi and T.~Berger, ``Rate--distortion for correlated sources with
  partially separated encoders,'' {\em IEEE Trans. Inform. Theory},
  vol.~\mbox{IT-28}, pp.~828--840, Nov. 1982.

\bibitem{SefidgaranT:12}
M.~Sefidgaran and A.~Tchamkerten, ``On cooperation in multi-terminal
  computation and rate distortion,'' in {\em Proc. IEEE Int. Symp. Inform.
  Theory}, (Cambridge, MA), pp.~771--775, July 2012.

\bibitem{NaI:11}
N.~Ma and P.~Ishwar, ``Some results on distributed source coding for
  interactive function computation,'' {\em IEEE Trans. Inform. Theory},
  vol.~57, pp.~6180--6195, Sept. 2011.

\bibitem{NaIG:12}
N.~Ma, P.~Ishwar, and P.~Gupta, ``Interactive source coding for function
  computation in collocated networks,'' {\em IEEE Trans. Inform. Theory},
  vol.~58, pp.~4289--4305, July 2012.

\bibitem{NitinawaratN:10}
S.~Nitinawarat and P.~Narayan, ``Perfect omniscience, perfect secrecy, and
  {S}teiner tree packing,'' {\em IEEE Trans. Inform. Theory}, vol.~56,
  pp.~6490--6500, Dec. 2010.

\bibitem{CourtadeW:10}
T.~Courtade and R.~Wesel, ``Efficient universal recovery in broadcast
  networks,'' in {\em Proc. 48th Ann. Allerton Conf. on Commun., Control and
  Comp.}, (Monticello, IL), pp.~1542--1549, Oct. 2010.

\bibitem{MartinianWZ:08}
E.~Martinian, G.~W. Wornell, and R.~Zamir, ``Source coding with distortion side
  information,'' {\em IEEE Trans. Inform. Theory}, vol.~54, pp.~4638--4665,
  Oct. 2008.

\bibitem{LinderZZ:99}
T.~Linder, R.~Zamir, and K.~Zeger, ``High-resolution source coding for
  non-difference distortion measures: {M}ultidimensional companding,'' {\em
  IEEE Trans. Inform. Theory}, vol.~45, pp.~548--561, Mar. 1999.

\bibitem{GershoG:92}
A.~Gersho and R.~M. Gray, {\em Vector Quantization and Signal Compression}.
\newblock Boston, MA: Kluwer Acad. Pub., 1992.

\bibitem{Neuhoff:93}
D.~L. Neuhoff, ``The other asymptotic theory of lossy source coding,'' in {\em
  Coding and Quantization} (R.~Calderbank, {G. D. Forney, Jr.}, and N.~Moayeri,
  eds.), vol.~14 of {\em DIMACS Series in Discrete Mathematics and Theoretical
  Computer Science}, pp.~55--65, American Mathematical Society, 1993.

\bibitem{GrayN:98}
R.~M. Gray and D.~L. Neuhoff, ``Quantization,'' {\em IEEE Trans. Inform.
  Theory}, vol.~44, pp.~2325--2383, Oct. 1998.

\bibitem{Bennett:48}
W.~R. Bennett, ``Spectra of quantized signals,'' {\em Bell Syst. Tech. J.},
  vol.~27, pp.~446--472, July 1948.

\bibitem{PanterD:51}
P.~F. Panter and W.~Dite, ``Quantizing distortion in pulse-count modulation
  with nonuniform spacing of levels,'' {\em Proc. IRE}, vol.~39, pp.~44--48,
  Jan. 1951.

\bibitem{BucklewW:82}
J.~A. Bucklew and G.~L. Wise, ``Multidimensional asymptotic quantization theory
  with $r$th power distortion measures,'' {\em IEEE Trans. Inform. Theory},
  vol.~IT-28, pp.~239--247, Mar. 1982.

\bibitem{CambanisG:83}
S.~Cambanis and N.~L. Gerr, ``A simple class of asymptotically optimal
  quantizers,'' {\em IEEE Trans. Inform. Theory}, vol.~IT-29, pp.~664--676,
  Sept. 1983.

\bibitem{Linder:91}
T.~Linder, ``On asymptotically optimal companding quantization,'' {\em Prob.
  Contr. Inform. Theory}, vol.~20, no.~6, pp.~475--484, 1991.

\bibitem{Goyal:00b}
V.~K. Goyal, ``High-rate transform coding: How high is high, and does it
  matter?,'' in {\em Proc. IEEE Int. Symp. Inform. Theory}, (Sorrento, Italy),
  p.~207, June 2000.

\bibitem{GrayG:77}
R.~M. Gray and A.~H. {Gray, Jr.}, ``Asymptotically optimal quantizers,'' {\em
  IEEE Trans. Inform. Theory}, vol.~IT-23, pp.~143--144, Feb. 1977.

\bibitem{Segall:76}
A.~Segall, ``Bit allocation and encoding for vector sources,'' {\em IEEE Trans.
  Inform. Theory}, vol.~IT-22, pp.~162--169, Mar. 1976.

\end{thebibliography}

\end{document}